\documentclass[10pt,twocolumn,letterpaper]{article}

\usepackage[ruled,vlined]{algorithm2e}
\usepackage[table]{xcolor}

\usepackage{cvpr}              % To produce the CAMERA-READY version

\definecolor{cvprblue}{rgb}{0.21,0.49,0.74}
\usepackage[breaklinks,colorlinks,allcolors=cvprblue]{hyperref}
\usepackage{bookmark}

\def\paperID{22668} % *** Enter the Paper ID here
\def\confName{CVPR}
\def\confYear{2026}

\title{DiffSoup: Direct Differentiable Rasterization of Triangle Soup\\
for Extreme Radiance Field Simplification}

\author{
Kenji Tojo$^{1}$ \quad
Bernd Bickel$^{2}$ \quad
Nobuyuki Umetani$^{1}$\\
$^{1}$The University of Tokyo \quad
$^{2}$ETH Zürich\\
{\tt\small \href{mailto:knjtojo@g.ecc.u-tokyo.ac.jp}{knjtojo@g.ecc.u-tokyo.ac.jp} \quad
\href{mailto:bickelb@ethz.ch}{bickelb@ethz.ch} \quad
\href{mailto:n.umetani@gmail.com}{n.umetani@gmail.com}}
}

\begin{document}

\maketitle
\begin{abstract}
Radiance field reconstruction aims to recover high-quality 3D representations from multi-view RGB images.
Recent advances, such as 3D Gaussian splatting, enable real-time rendering with high visual fidelity on sufficiently powerful graphics hardware.
However, efficient online transmission and rendering across diverse platforms requires drastic model simplification, reducing the number of primitives by several orders of magnitude.
We introduce DiffSoup, a radiance field representation that employs a soup (i.e., a highly unstructured set) of a small number of triangles with neural textures and binary opacity.
We show that this binary opacity representation is directly differentiable via stochastic opacity masking, enabling stable training without a mollifier (i.e., smooth rasterization).
DiffSoup can be rasterized using standard depth testing, enabling seamless integration into traditional graphics pipelines and interactive rendering on consumer-grade laptops and mobile devices.
Code is available at \url{https://github.com/kenji-tojo/diffsoup}.%
\end{abstract}

\section{Introduction}
\label{sec:intro}
Realistic and efficient 3D representations are essential for a wide range of vision and graphics applications, including virtual and augmented reality, digital content creation, and robotic perception.
However, it remains challenging to represent complex real-world objects such as foliage, clothing, hair, or architectural details across diverse platforms, including mobile and VR devices.
Typically, the number of geometric primitives (e.g., triangles) is tightly limited for interactive rendering and smooth data transfer.

Neural radiance fields (NeRFs)~\cite{mildenhall2020nerf} have demonstrated high-fidelity 3D reconstruction from multi-view RGB images via differentiable volume rendering~\cite{kajiya1984ray}.
More recently, 3D Gaussian Splatting (3DGS)~\cite{kerbl3Dgaussians} models the radiance field as point-based primitives, enabling real-time rendering and flexible scene editing given sufficiently powerful graphics hardware.
Although these approaches—and many subsequent variants—achieve remarkable view synthesis results, they typically rely on an enormous number of primitives, often \emph{orders of magnitude} more than the number feasible on restricted platforms and devices.

Traditional graphics pipelines are highly optimized for rendering opaque triangles with texture maps.
However, opaque triangles introduce difficulties in training: unlike volumetric primitives, they fully occlude surfaces behind them and produce discontinuous color transitions at silhouettes, both of which impede smooth gradient propagation.
Consequently, differentiable rasterization methods for opaque triangles~\cite{Laine2020diffrast,pidhorskyi2024rasterized} are still not widely adopted in radiance field reconstruction pipelines.

\begin{figure}
\centering
\includegraphics[width=\linewidth]{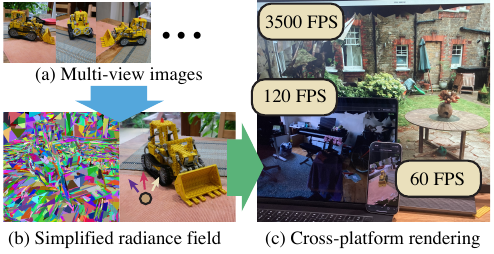}
\caption{%
(a) Given multi-view RGB images, (b) we reconstruct a simplified radiance field as a textured triangle soup using differentiable rasterization.
(c) The resulting scene can be rendered within a standard depth-tested rasterization pipeline, enabling seamless visualization across devices—including a high-end desktop with a dedicated GPU, a consumer-grade laptop (MacBook Pro with an M3 Pro chip), and even a smartphone (iPhone 15).
}
\label{fig:teaser}
\end{figure}

In this paper, we address the challenging problem of \emph{extreme model simplification}~\cite{decoret2003billboard} for radiance field–based view synthesis.
Our goal is to reconstruct a 3D scene from multi-view RGB images using a small unstructured collection (or \emph{soup}) of opaque triangles equipped with compact neural textures that have binary opacity.
The resulting representation seamlessly integrates into traditional real-time rendering pipelines while preserving the optimization robustness and reconstruction fidelity typically associated with volumetric approaches.

To achieve this, we introduce \emph{DiffSoup}, a differentiable rasterization method for directly learning an opaque triangle soup.
Unlike previous approaches, DiffSoup does not rely on a \emph{molifier} (i.e., smoothed rasterization) during optimization, which often suppresses high-frequency appearance details or requires careful scheduling of blur strength~\cite{Held2025Triangle,liu2019softras}.
Instead, inspired by recent work on implicit surface reconstruction~\cite{Zhang2025Radiance}, we differentiate opaque triangle rasterization using \emph{stochastic opacity masking}, which yields robust gradient signals under severe occlusion while producing the \emph{exact} pixel color.
To support this, DiffSoup further extends the silhouette-differentiation method of~\cite{pidhorskyi2024rasterized} to handle the \emph{implicit} discontinuities introduced by opacity masking.

In summary, our contributions are as follows.

\begin{itemize}
\item We introduce the \emph{extreme radiance field simplification} problem, aiming to represent a 3D scene using an unstructured soup of a small number of textured triangles.

\item We present \emph{DiffSoup}, a method for directly differentiating opaque triangle rasterization via \emph{stochastic opacity masking}, yielding robust gradients even under severe occlusion and frequent visibility discontinuities.

\item We demonstrate that DiffSoup reconstructs high-fidelity scenes using typically fewer than 20,000 triangles—where existing methods rely on millions of primitives—producing more detail-preserving novel views than baselines, while running interactively on consumer-grade laptops and mobile phones (Figure~\ref{fig:teaser}).
\end{itemize}

\begin{figure*}
\centering
\includegraphics[width=\linewidth]{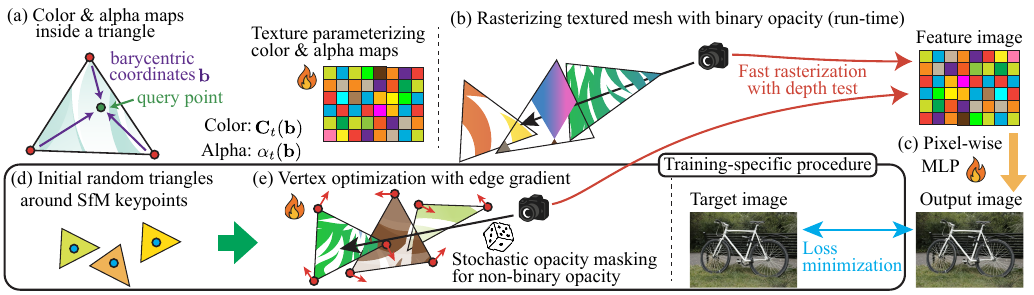}
\caption{%
Overview of our algorithm.
(a) We represent radiance fields as a textured triangle soup, where barycentric coordinates map to learnable color and alpha textures.
(b) At run time, this representation is rendered with binary opacity, enabling efficient rasterization via standard depth testing.
(c) Rasterization produces a high-dimensional color-feature image, which is then mapped to RGB values using a small shared MLP trained to minimize the error between the rendering and the ground-truth image.
(d) During training, we initialize triangles around structure-from-motion (SfM) keypoints.
(e) We optimize opacity through stochastic opacity masking, while vertex positions are updated using edge gradients for opaque triangles.
}
\label{fig:overview}
\end{figure*}

\section{Related Work}
\label{sec:related}

\paragraph{Differentiable rendering}
A large body of work reconstructs 3D scenes from images via differentiable rendering (see survey~\cite{kato2020differentiable}).
These methods optimize geometry, texture, and material parameters using gradients from rasterization~\cite{liu2019softras,Laine2020diffrast,ravi2020accelerating,loper2014opendr,Hasselgren2021}, light transport simulation~\cite{NimierDavidVicini2019Mitsuba2,Li:2018:DMC,Loubet2019Reparameterizing,zhangmanyworld}, implicit surface rendering~\cite{Bangaru2022NeuralSDFReparam,Vicini2022sdf,wang2021neus,li2023neuralangelo,Zhang2025Radiance,yariv2021volume,Mehta_2023_ICCV}, multi-layer transparency models~\cite{Hasselgren2021,Esposito2025VolSurfs}, point splatting~\cite{yifan2019differentiable,kerbl3Dgaussians}, and more.
Our work focuses on opaque triangle rasterization with a single depth-tested layer, the dominant rendering method in real-time applications such as games, VR/AR, and mobile graphics.

\paragraph{Differentiable triangle rasterization}
Computing gradients for opaque, depth-tested rasterization is challenging due to occlusion and the discontinuities that arise at object silhouettes.
Soft-rasterization methods introduce customized edge smoothing to enforce differentiability~\cite{liu2019softras,Held2025Triangle,Held2025Triangle+}, while other approaches derive gradients directly from anti-aliasing formulations to preserve sharp silhouettes~\cite{loper2014opendr,chen2019dibrender,Laine2020diffrast,pidhorskyi2024rasterized}.
Although these ``blur-free'' methods better preserve high-frequency appearance details, their gradients are often unstable and non-smooth, leading to suboptimal reconstruction~\cite{Nicolet2021Large}.
To address this, we draw inspiration from the \emph{stochastic surface} interpretation developed for implicit surface reconstruction~\cite{Zhang2025Radiance} and inverse light transport~\cite{zhangmanyworld}.
Specifically, we extend differentiable rasterizers~\cite{Laine2020diffrast,pidhorskyi2024rasterized} with \emph{stochastic opacity masking}, which yields robust gradients for optimizing binary opacity under occlusion while still differentiating the exact pixel colors produced by opaque rasterization.

\paragraph*{Neural radiance fields}
Radiance-field reconstruction methods such as NeRF and its extensions~\cite{mildenhall2020nerf,barron2022mipnerf360,mueller2022instant,plenoxels,barron2021mipnerf,adaptiveshells2023} recover high-quality 3D representations from multi-view RGB images.
Despite their strong view-synthesis quality, integrating these representations into traditional real-time rendering pipelines remains difficult due to costly volumetric ray marching~\cite{kajiya1984ray}.
MobileNeRF~\cite{chen2022mobilenerf} converts a radiance field into a mesh-based representation but does not directly differentiate the final opaque mesh, relying instead on NeRF optimization as a separate intermediate stage. This limits control over polygon count and final geometry, often producing excessive triangles and losing high-frequency appearance details.
In contrast, we extend direct differentiable rasterization for polygonal asset creation~\cite{pidhorskyi2024rasterized,Laine2020diffrast,Hasselgren2021} to radiance-field reconstruction, enabling significantly more compact, detail-preserving results with shorter training.

\paragraph*{3D Gaussian splatting}
The 3DGS method uses explicit 3D Gaussian primitives for real-time radiance-field rendering~\cite{kerbl3Dgaussians}.
To capture more accurate geometry and sharper details, follow-up works explore alternative primitives such as 2D Gaussians~\cite{Huang2DGS2024}, convex hulls~\cite{Held20243DConvex}, triangles~\cite{Held2025Triangle}, and textured 2D~\cite{xu2024SuperGaussians,xu2024texturegs,rong2024gstex,svitov2024billboard,weiss2024gaussian} or 3D~\cite{chao2025texturedgaussians,xu2024texturegs} primitives.
However, these approaches often rely on excessive primitive counts or introduce custom translucency, both of which limit practicality on consumer-grade devices such as laptops, mobile phones, and VR headsets.
Concurrent work~\cite{Held2025Triangle+} proposes a scheduling scheme for the smoothing function in triangle rasterization and shows that scenes can converge to opaque triangles.
Their method, however, does not optimize sub-triangle appearance using opacity and texture, and therefore requires many triangles to represent fine details.
In contrast, we directly optimize an opaque triangle soup with a neural texture carrying binary opacity, enabling high-frequency details with a drastically smaller number of primitives.
StochasticSplats~\cite{kv2025stochasticsplats} applies a stochastic differentiation strategy to optimize the \emph{translucent} appearance of 3DGS using multiple random samples; we instead target the \emph{fully opaque} appearance of triangle rasterization using a single depth-tested stochastic rasterization.

\section{Method}
\label{sec:method}

Given multi-view RGB images, DiffSoup reconstructs a triangle soup augmented with neural textures and binary opacity.
We first outline the necessary background and then describe our differentiable rasterizer, along with the proposed stochastic opacity masking approach.
An overview of our algorithm is shown in Figure~\ref{fig:overview}.

\subsection{Preliminary}
\paragraph*{Triangle soup representation}
To represent a radiance field, we use triangles $\mathcal{T}=\{\mathbf{T}_t\}_{t=1}^{\vert\mathcal{T}\vert}$ together with per-triangle texture functions for color $\mathcal{C}=\{\mathbf{C}_t\}_{t=1}^{\vert\mathcal{T}\vert}$ and opacity $\mathcal{A}=\{\alpha_t\}_{t=1}^{\vert\mathcal{T}\vert}$. 
Each triangle $\mathbf{T}_t$ consists of three vertices $\mathbf{v}_1,\mathbf{v}_2,\mathbf{v}_3\in\mathbb{R}^3$. 
Given an appropriate parameterization, the color texture $\mathbf{C}_t$ maps barycentric coordinates $\mathbf{b}\in\mathbb{R}^2$ to the color $\mathbf{C}_t(\mathbf{b})$ at the corresponding point on the triangle, while opacity is defined by a scalar function $\alpha_t(\mathbf{b})\in[0,1]$.

\paragraph*{Depth-tested rasterization with binary opacity}
Unlike previous approaches that rely on translucent primitives, DiffSoup employs opaque triangle rasterization with binary opacity, enabling seamless integration into traditional graphics pipelines.
Given triangles transformed into camera space, we perform standard rasterization to produce an array of fragments $\mathcal{F}=\{\mathbf{f}_f\}_{f}^{\vert\mathcal{F}\vert}$, defined for each individual pixel. 
Fragments are generated for all triangles intersecting the pixel and encode the pixel-center sample as
\begin{equation}
    \mathbf{f}_f = (\mathbf{b}_f, D_f, t_f),
\end{equation}
where $\mathbf{b}_f$ denotes the barycentric coordinates, $D_f$ is the fragment depth, and $t_f$ is the corresponding triangle index.
Given this information, the pixel color $\hat{\mathbf{C}}$ is computed as
\begin{equation}
    \hat{\mathbf{C}} = \mathbf{C}_{t_{f^*}}(\mathbf{b}_{f^{*}}),
\end{equation}
where $f^{*}=\arg\min_{f} D_f$ is the index of the \emph{foremost} fragment.
Hereafter, we write $\mathbf{C}_f$ and $\alpha_f$ as shorthands for $\mathbf{C}_{t_f}$ and $\alpha_{t_f}$, respectively, to simplify notation.

In practical real-time rendering systems such as games, sub-triangle geometric detail is commonly represented using \emph{binary opacity} stored in a texture.
This modifies the selection of the foremost fragment by discarding the fragments whose opacity falls below a threshold:
\begin{equation}
\label{eq:binary}
    f^{*}=\underset{f\in\{f;\,\alpha(\mathbf{b}_f)\,>\,0.5\}}{\arg\min}\,
    D_f.
\end{equation}
This enables fine sub-triangle detail using a small number of flat primitives, without adding complexity beyond depth-tested rendering, such as the explicit primitive sorting used in 3DGS~\cite{kerbl3Dgaussians}.

\paragraph*{Stochastic surface and radiance field loss}
Although forward rendering with opaque rasterization and binary opacity is highly efficient, \emph{directly optimizing} such representations remains difficult because occlusion and visibility transitions are discontinuous.
Prior work~\cite{chen2022mobilenerf} employs a straight-through estimator~\cite{bengio2013estimating} to approximate gradients with respect to binarized opacity values during NeRF training.
However, this strategy exhibits instability in geometry optimization and requires an additional pre-training stage using a standard NeRF.

Recent work has introduced an important advancement for implicit surface reconstruction~\cite{Zhang2025Radiance,zhangmanyworld,Mehta_2023_ICCV}.
In particular, Zhang et al.~\cite{Zhang2025Radiance} interpret \emph{every} point in space as an infinitesimal opaque surface occurring with probability equal to that point's opacity.
This \emph{stochastic surface} interpretation leads to a modified NeRF training loss:
\begin{equation}
\label{eq:radiance}
    \mathcal{L}_{\mathrm{sum}}
    =
    \sum_{i}\,
    w_i\,
    \mathcal{L}_{1}(\mathbf{C}_i),
    \quad
    \text{where }
    w_i
    =
    \bar{\alpha}_i\,
    \alpha_i.
\end{equation}
Here, $\mathbf{C}_i$ and $\alpha_i$ denote the color and alpha values at the $i$-th sample point along the ray in depth-sorted order, and $\bar{\alpha}_i=\Pi_{j=1}^{i-1}(1-\alpha_j)$ is the cumulative transmittance.
Notably, they evaluate the pixel-wise L1 loss $\mathcal{L}_{1}$ at the \emph{sampled} points with weights proportional to the volumetric blending weights, rather than computing $\mathcal{L}_{1}$ on the final blended pixel color as in standard NeRFs.
They show that this simple, theoretically grounded loss converges to binary opacity and exactly reproduces pixel colors in opaque rendering, without heuristics such as smoothness scheduling or pre-training.

However, their method seeks a potential surface via sampling in \emph{continuous} space, which is incompatible with the \emph{inherently discrete} rasterization pipeline, where fragments arise from explicit primitives and depth sorting introduces substantial complexity.
Moreover, their formulation omits gradients with respect to the motion of explicit primitives; as a result, when applied within a triangle-based rasterization pipeline, it yields markedly suboptimal geometry.

\subsection{The DiffSoup differentiable rasterizer}
Our goal is to inversely optimize the previously introduced triangle-soup representation via depth-tested rasterization, given ground-truth multi-view RGB images.
To this end, we build on differentiable rasterization methods for opaque triangles~\cite{Laine2020diffrast,pidhorskyi2024rasterized}, which derive gradients by differentiating an anti-aliasing model.
While these methods provide a solid foundation for differentiating visibility discontinuities while preserving sharp edges, they struggle to optimize binary opacity and to handle \emph{implicit} visibility discontinuities—those not associated with explicit triangle edges—introduced by opacity masking.

\begin{figure}[t]
\centering
\includegraphics[width=\linewidth]{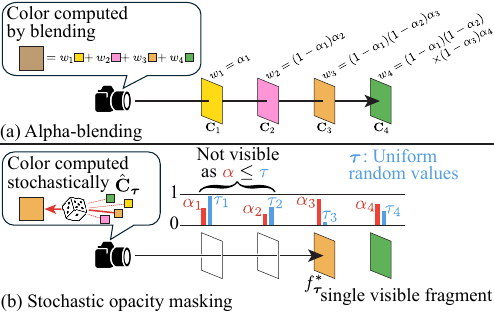}
\caption{%
(a) Alpha blending computes cumulative transmittance over depth-sorted samples and produces the final color as a weighted average.
(b) Our stochastic opacity masking, in contrast, simulates a discrete stochastic process over the surface, allowing the renderer to output the exact color of the selected fragment.
}
\label{fig:stochastic}
\end{figure}

\paragraph*{Stochastic opacity masking}
To derive a differentiable optimization method, we extend the stochastic surface approach~\cite{Zhang2025Radiance} to our differentiable triangle rasterizer.
However, a standard object-order rasterizer does not produce depth-sorted fragments per pixel, so the radiance field loss~\eqref{eq:radiance} cannot be directly evaluated.
We therefore introduce \emph{stochastic opacity masking}, a principled mechanism that realizes the stochastic surface process entirely within a conventional depth-tested rasterization pipeline.

The key idea is to \emph{simulate} the stochastic surface as a \emph{discrete} stochastic process during fragment generation (Figure~\ref{fig:stochastic}).
Algorithmically, DiffSoup achieves this by introducing stochastic opacity thresholds into~\eqref{eq:binary} as
\begin{equation}
\label{eq:stochastic}
    f^{*}_{\boldsymbol{\tau}}
    =
    \underset{\bar{f}\in\{f\in\mathcal{F}|\,\alpha(\mathbf{b}_f)\,>\,\tau_f\}}{\arg\min}\,
    D_{\bar{f}},
\end{equation}
where $\boldsymbol{\tau}=(\tau_1,\,\dots,\,\tau_{\vert\mathcal{F}\vert})$ are real numbers sampled independently and uniformly for each fragment in the range $[0,1]$.
This selects a single visible surface point for the pixel, and the resulting color is $\hat{\mathbf{C}}_{\boldsymbol{\tau}} = \mathbf{C}_{f^{*}_{\boldsymbol{\tau}}}(\mathbf{b}_{f^{*}_{\boldsymbol{\tau}}})$.
Note that the $\arg\min$ in \eqref{eq:stochastic} can be efficiently computed using the z-buffer method without depth sorting the fragments.

The crucial observation is that stochastic masking makes the pixel color a \emph{random variable}: high-opacity fragments are more likely to pass the threshold and become visible, though even the most plausible surface may be skipped with a small probability.
Consequently,~\eqref{eq:stochastic} can be interpreted as drawing a single pixel color from the probability distribution induced by the stochastic surface.
Specifically, for a scene parameter $\theta$, we can show that the probability that fragment $f$ is selected as the foremost surface is identical to the weight in~\eqref{eq:radiance} as 
\begin{equation}
    p(f;\theta)=w_f,
\end{equation}
whereas our computation does not require sorting fragments by depth.
With this identity, the radiance field loss~\eqref{eq:radiance} is computed as the \emph{expected value} of the pixel-wise loss:
\begin{equation}
\label{eq:exp}
    \mathcal{L}_{\mathrm{exp}}
    =
    E_{p(f;\theta)}
    [
    \mathcal{L}_{1}(\hat{\mathbf{C}})
    ],
\end{equation}
as estimated by our stochastic opacity masking.

Notably, viewing the rasterization as a discrete stochastic process enables unbiased estimation of the gradient of $\mathcal{L}_{\mathrm{exp}}$ using the DiffSoup rasterizer as a sampler.
Specifically, we apply the likelihood-ratio gradient identity~\cite{williams1992simple,glynn1990likelihood}:
\begin{equation}
\begin{aligned}
    \partial_{\theta} E_{p(f;\theta)}[\mathcal{L}_1(\hat{\mathbf{C}})]
    &=
    E_{p(f;\theta)}[\partial_{\theta}\mathcal{L}_1(\hat{\mathbf{C}})] \\
    &+
    E_{p(f;\theta)}[
        \mathcal{L}_1(\hat{\mathbf{C}})\,
        \partial_{\theta}\log p(f;\theta)
    ],
\end{aligned}
\end{equation}
where $\partial_{\theta}$ denotes the partial derivative with respect to the scene parameter $\theta$.
The first term is the standard loss gradient for the \emph{exact} color of the sampled fragment.
The second term propagates gradients with respect to opacity, where the score term $\partial_{\theta}\log p(f;\theta)$ is weighted by the pixel-wise color loss.
For a fragment $f'$, this score term reduces to the partial derivative with respect to its opacity:
\begin{equation}
\frac{\partial}{\partial\alpha_{f'}}\log p(f;\theta)
=
\begin{cases}
    1/\alpha_f, & f'=f,\\
    -1/(1-\alpha_{f'}), & D_{f'} < D_f,
\end{cases}
\end{equation}
and no gradient is produced for fragments with $D_{f'} > D_f$.
Crucially, the opacity gradient for each fragment $f'$ depends only on the sampled pixel fragment $f$ and the fragment itself, yielding an efficient, sort-free gradient estimator that operates entirely within the depth-tested rasterizer (pseudocode in the supplementary material).

\paragraph*{Handling visibility discontinuities}
Stochastic opacity masking yields gradients for binary opacity but not for the \emph{motion} of explicit primitives.
The approach of~\cite{Zhang2025Radiance}, which relies on dense spatial sampling for implicit fields, is incompatible with our primitive-based setting. To obtain motion gradients, we extend the \emph{edge-gradient} formulation of differentiable rasterization~\cite{pidhorskyi2024rasterized}.
Conventional edge gradients detect visibility boundaries by examining changes in triangle IDs between neighboring pixels.
In DiffSoup, however, stochastic opacity masking introduces \emph{intra-triangle} visibility discontinuities that do not alter triangle IDs and therefore remain undetected.
To handle these internal discontinuities, we evaluate the rasterized edge-gradient expression for \emph{all} horizontally and vertically adjacent pixel pairs in the stochastic intermediate image when computing the color loss.
This naturally generalizes edge gradients to implicit, opacity-induced boundaries.
Averaged over training iterations, this procedure produces stable and meaningful motion gradients that align with the effective sub-triangle silhouettes emerging from converged binary opacity, enabling accurate optimization of triangle geometry and placement.

\begin{figure}[t]
\centering
\includegraphics[width=\linewidth]{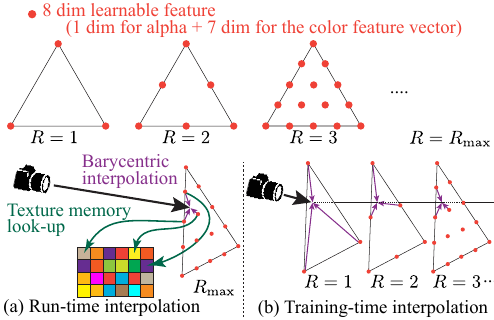}
\caption{%
Multi-resolution triangle texture.
Top: We parameterize the texture using multiple grid resolutions, corresponding to different triangle subdivision levels.
Bottom: (a) At run time, the renderer queries only the highest-resolution triangle texture. (b) At training time, this highest-resolution texture is over-parameterized with lower-resolution grids to improve optimization stability.
}
\label{fig:multires}
\end{figure}

\definecolor{BestCell}{HTML}{DFF0D8}
\definecolor{SecondCell}{HTML}{D9EDF7}

\begin{figure*}
\centering
\includegraphics[width=\linewidth]{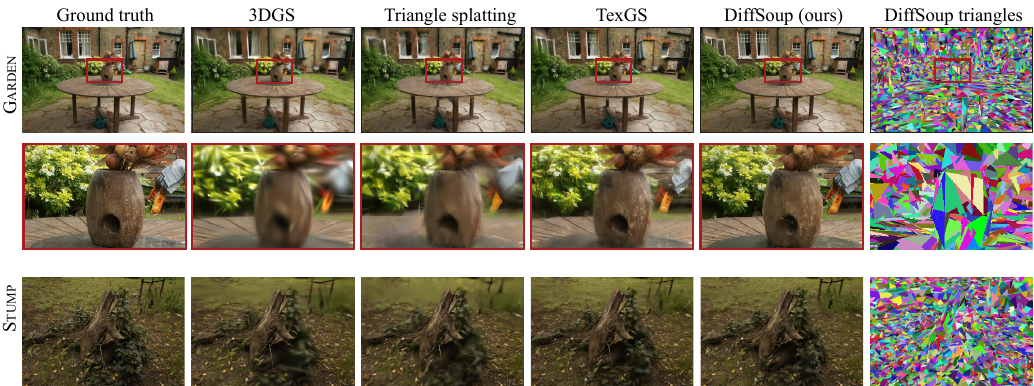}
\caption{%
Qualitative results on the \textit{MipNeRF360}~\cite{barron2022mipnerf360} dataset.
(Left to right) Ground-truth test images; views synthesized by 3DGS~\cite{kerbl3Dgaussians}, Triangle Splatting~\cite{Held2025Triangle}, TexGS~\cite{chao2025texturedgaussians}, our method; and a visualization of our reconstructed triangle geometry.
Rows correspond to different scenes, and all models use 15K primitives.
}
\label{fig:mip360}
\end{figure*}

\subsection{Parameterization}
\paragraph*{Learnable multi-resolution triangle texture}
A straightforward way to represent per-triangle textures $\mathbf{C}_t$ and $\alpha_t$ is to map the triangle soup onto a 2D texture atlas and store texture values directly in image space.
However, such an atlas may place unrelated triangles adjacent in texture space, which interferes with the use of \emph{multi-resolution} parameterizations that have proven highly effective in gradient-based optimization, such as multi-level 3D grids~\cite{mueller2022instant,takikawa2021nglod} and texture mip-mapping~\cite{Weier2025PracticalInverse,Laine2020diffrast}.
To avoid these issues and to fully leverage multi-resolution optimization, we adapt these parameterizations to operate directly on our triangle soup.

Specifically, we extend the ``mesh color'' approach~\cite{yuksel2010mesh,yuksel2017mesh} for atlas-free texturing to support \emph{learnable} feature vectors.
At each resolution level $R$, we place learnable features at the vertices of a recursively subdivided triangle grid: the base level $R=0$ corresponds to the original triangle vertices, and each subsequent level $R$ refines the previous one, yielding $(2^{R-1}+1)(2^{R}+1)$ feature vertices (Figure~\ref{fig:multires}).
These multi-resolution per-triangle features are queried via barycentric interpolation within the micro-triangle containing the evaluation point.

To enable a multi-resolution representation of the texture at target level $R_{\mathrm{max}}$, we maintain learnable feature grids for $R_{\mathrm{min}},\,\dots,\,R_{\mathrm{max}}-1$.
All lower-resolution grids are queried and accumulated on the finest grid before rendering, yielding an \emph{overparameterized} high-resolution texture field, similar in spirit to~\cite{Weier2025PracticalInverse} but applied to a triangular grid rather than a texture image.
A sigmoid activation is applied to constrain the resulting features to $[0,1]$ prior to rasterization.
This strategy preserves the benefits of multi-resolution optimization while avoiding multiple per-level texture lookups during rendering.
After training, only the finest-resolution texture (after sigmoid activation) is retained and stored as 8-bit PNG images.

\paragraph*{Neural deferred shading}
To model view-dependent appearance without increasing texture storage or relying on high-order bases (e.g., spherical harmonics), we employ neural deferred shading~\cite{thies2019deferred,worchel:2022:nds} within our binary-opacity rasterization framework.
Rather than storing RGB values directly, each triangle carries an $N_{\mathrm{feat}}$-dimensional learnable feature vector in its color texture, which is rasterized to yield per-pixel feature maps.
A lightweight shared MLP is then applied per pixel, taking as input the concatenation of the rasterized feature vector and the viewing direction, and producing the final shaded color.
Opacity, on the contrary, is stored as a scalar texture and kept \emph{outside} the neural network.
This separation reflects its view-independent nature and ensures efficient access during stochastic opacity masking, where opacity must be evaluated for \emph{all} fragments, while only the ``winning’’ fragment’s color participates in differentiation.
We use $N_{\mathrm{feat}}=7$, allowing all features and opacity to fit in two 8-bit RGBA textures after training.

\subsection{Optimization}

\paragraph*{Loss function}
To optimize our triangle soup representation, we minimize a photometric objective,
\begin{equation}
    \lambda\,\mathcal{L}_{1} + (1-\lambda)\,\mathcal{L}_{\mathrm{SSIM}},
\end{equation}
where $\mathcal{L}_{\mathrm{SSIM}}$ denotes the structural similarity term~\cite{wang2004image}.
We set $\lambda = 0.8$ following 3DGS~\cite{kerbl3Dgaussians}.
The loss is evaluated on the stochastic rendering, analogous to~\eqref{eq:exp}.

\paragraph*{Initialization and training details}
To better capture high-frequency details through texture, we sample two-thirds of the target points using farthest point sampling (FPS)~\cite{eldar1997farthest} on the SfM points and draw the remaining one-third uniformly at random.
We then randomly initialize triangles around these sampled points with a radius set to one quarter of the average nearest-neighbor distance.
We use VectorAdam~\cite{ling2022vectoradam} for optimizing vertex positions and  Adam~\cite{kingma2014adam} for all other parameters.
Training is run for 10{,}000 iterations.
Additional hyperparameters are provided in the supplementary material.
We compute gradients for four views in parallel at each optimization step.

\paragraph*{Coarse-to-fine optimization}
We find that coarse-to-fine texture optimization improves the convergence of both geometry and color.
Specifically, we use $R_{\mathrm{min}} = R_{\mathrm{max}} = 3$ (i.e., vertex features) for the first $5000$ iterations, and switch to $R_{\mathrm{min}} = 2$ and $R_{\mathrm{max}} = 5$ for the final $5000$ iterations.

\paragraph*{Adaptive control of primitives}
To balance reconstruction fidelity with a compact primitive budget, we introduce an adaptive mechanism for dynamically adjusting the triangle set.
To avoid overly large primitives that compromise high-frequency detail, we periodically subdivide edges whose view-space length exceeds a prescribed fraction of the image height (1/5 in our experiments).
This encourages approximately uniform view-space edge lengths and improves the effective use of texture resolution.
Conversely, we enforce the target primitive count by pruning triangles with low pixel coverage across the training views until the desired budget is met.
In practice, edge splitting and triangle removal are performed every $100$ iterations.
For edge splitting, view-space edge lengths are computed on a randomly sampled subset of $20$ training views, as evaluating all views would introduce significant overhead.

\begin{table}
\caption{
Quantitative evaluation on the \textit{MipNeRF360}~\cite{barron2022mipnerf360} dataset.
For PSNR, SSIM, and LPIPS, we report scene-level averages. The number of primitives is strictly enforced for all methods.
}
\label{tab:mip360}
\centering
\small
\begin{tabular}{lcccc}
\toprule
Method & PSNR $\uparrow$ & SSIM $\uparrow$ & LPIPS $\downarrow$ & \# Prims \\
\midrule
3DGS~\cite{kerbl3Dgaussians} & 23.72 & 0.664 & 0.420 & 15K \\ 
TS~\cite{Held2025Triangle} & 22.81 & 0.634 & 0.430 & 15K \\ 
TexGS~\cite{chao2025texturedgaussians} & \cellcolor{BestCell}\textbf{24.80} & \cellcolor{SecondCell}0.697 & \cellcolor{SecondCell}0.270 & 15K \\ 
\textbf{Ours} & \cellcolor{SecondCell}24.76 & \cellcolor{BestCell}\textbf{0.748} & \cellcolor{BestCell}\textbf{0.204} & 15K \\ 
\bottomrule
\end{tabular}
\end{table}

\begin{figure*}
\centering
\includegraphics[width=\linewidth]{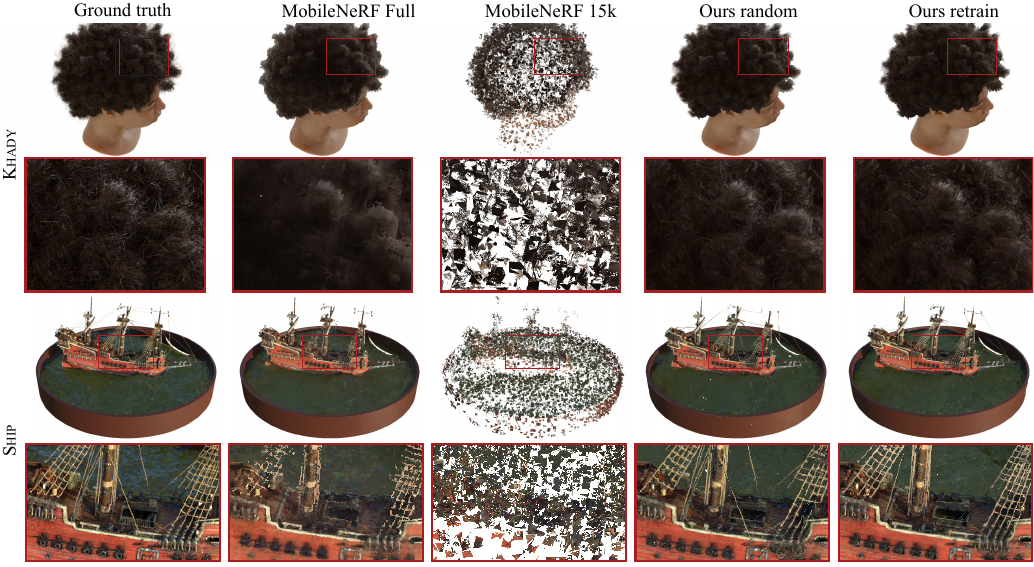}
\caption{%
Qualitative results on synthetic scenes.
(Left to right) Ground-truth views; view-synthesis results from MobileNeRF; a mesh decimated from the MobileNeRF output using QEM~\cite{garland1997}; our results optimized from random initialization; and our results initialized from the MobileNeRF-decimated mesh.
}
\label{fig:synthetic}
\end{figure*}

\begin{table*}
\caption{
Comparison with MobileNeRF~\cite{chen2022mobilenerf} on synthetic scenes from the \textit{NeRF-Synthetic}~\cite{mildenhall2020nerf} and \textit{Shelly}~\cite{adaptiveshells2023} datasets.
We report PSNR for each scene. The face count for MobileNeRF is averaged across scenes, whereas all other methods use exactly 15K faces.
}
\label{tab:synthetic}
\centering
\small
\begin{tabular}{lccccccccc}
  \toprule
  Method & \multicolumn{5}{c}{\textit{NeRF-Synthetic}~\cite{mildenhall2020nerf} (PSNR $\uparrow$)} & \multicolumn{4}{c}{\textit{Shelly}~\cite{adaptiveshells2023} (PSNR $\uparrow$)} \\
  \cmidrule(lr){2-6}
  \cmidrule(lr){7-10}
   & \textsc{ship} & \textsc{chair} & \textsc{mic} & \textsc{drums} & \# Faces $\downarrow$ & \textsc{khady} & \textsc{kitten} & \textsc{pug} & \# Faces $\downarrow$ \\
  \midrule
  MobileNeRF~\cite{chen2022mobilenerf} & \cellcolor{SecondCell}26.06 & 31.02 & 29.18 & \cellcolor{SecondCell}23.82 & 159K & 26.22 & 30.05 & 29.98 & 275K \\
  MobileNeRF + QEM~\cite{garland1997} & 8.44 & 16.34 & 17.35 & 13.35 & 15K & 11.21 & 14.42 & 11.25 & 15K \\
  \midrule
  Ours w/ QEM initialization & \cellcolor{BestCell}\textbf{26.68} & \cellcolor{BestCell}\textbf{32.11} & \cellcolor{BestCell}\textbf{30.18} & \cellcolor{BestCell}\textbf{23.94} & 15K & \cellcolor{SecondCell}26.55 & \cellcolor{BestCell}\textbf{31.23} & \cellcolor{BestCell}\textbf{30.51} & 15K \\
  Ours w/ random initialization & 25.29 & \cellcolor{SecondCell}31.71 & \cellcolor{SecondCell}29.64 & 23.47 & 15K & \cellcolor{BestCell}\textbf{26.67} & \cellcolor{SecondCell}30.68 & \cellcolor{SecondCell}30.16 & 15K \\
  \bottomrule
\end{tabular}
\end{table*}

\section{Experiments}
\label{sec:experiments}

\paragraph*{Real-world dataset}
We evaluate our method on novel-view synthesis from multi-view RGB inputs using the MipNeRF360 dataset~\cite{barron2022mipnerf360}.
We compare against state-of-the-art primitive-based radiance-field reconstruction approaches.
As shown in the qualitative results in Figure~\ref{fig:mip360}, 3DGS~\cite{kerbl3Dgaussians} is not well suited for low primitive budgets (e.g., 15K) and struggles to capture fine details and sharp boundaries. Because the method provides no explicit mechanism for controlling the final primitive count, we enforced a fixed budget by sub-sampling the initial SfM points and disabling adaptive densification.

TriangleSplatting~\cite{Held2025Triangle} (TS) employs triangular primitives with a Markov chain Monte Carlo formulation~\cite{kheradmand20243d} to control the primitive count, but representing complex scenes without texture in this low-poly regime remains fundamentally challenging.
TexturedGaussians~\cite{chao2025texturedgaussians} (TexGS) introduces texture fields to encode high-frequency appearance without increasing the primitive count. While TexGS significantly enhances detail reconstruction under constrained budgets, it still struggles to reproduce crisp, well-defined boundaries.

In contrast, our stochastic opacity masking drives primitive opacities toward binary values, enabling the reconstruction of the sharpest boundaries and the most detail-preserving view-synthesis results among all methods. These observations are further corroborated by the quantitative metrics in Table~\ref{tab:mip360}.

\paragraph*{Rendering efficiency}
Beyond reducing primitive count, our approach enables hardware-accelerated rasterization by eliminating the transparency handling and depth sorting required in prior work.
Table~\ref{tab:fps_mip360} reports average rendering cost on an NVIDIA RTX~4090.
Although 3DGS benefits from a highly optimized CUDA renderer, our hardware-accelerated triangle rasterization achieves an order-of-magnitude speedup, despite incorporating additional texture information.
Moreover, by relying on standard graphics pipelines, the same shader runs on commodity devices (e.g., the MacBook in Figure~\ref{fig:teaser}), yielding performance competitive with CUDA-based 3DGS on a desktop GPU.
We fix primitive count across all methods, as it primarily governs the depth-sorting cost that dominates rendering time.
For instance, TexGS uses $170\times$ more tunable parameters than 3DGS yet is only about $10\times$ slower, indicating that adding detail through texture is more efficient than increasing primitive count.

\paragraph*{Synthetic scenes}
We further evaluate our method on synthetic scenes from the \textit{NeRF-Synthetic}~\cite{mildenhall2020nerf} and \textit{Shelly}~\cite{adaptiveshells2023} datasets.
In this setting, we compare our approach with MobileNeRF~\cite{chen2022mobilenerf}, which converts a trained NeRF~\cite{mildenhall2020nerf} into a triangle mesh with a binary opacity texture.
Although MobileNeRF can reconstruct fine structures with complex topology, it typically produces an excessively large number of faces per object (often exceeding 150K). Adjusting its training configuration to control the final face count is non-trivial, as training is computationally expensive (over 6 hours in our experiments).
Due to MobileNeRF’s grid-aligned mesh topology, standard mesh simplification algorithms—such as quadratic error metric (QEM) decimation~\cite{garland1997}—are ineffective at reducing the face count while preserving detail.

Our method naturally supports such \emph{geometric simplification} scenarios when a coarse geometric prior is available.
For instance, as shown in Figure~\ref{fig:synthetic}, we can uniformly sample triangle primitives within the bounding box of the MobileNeRF representation and iteratively prune transparent primitives to obtain a clean triangle soup. Alternatively, we can use their low-quality decimated mesh as an \emph{initialization} for our optimization.
In both cases, our method achieves comparable reconstruction fidelity with a drastically reduced triangle count and significantly less time than retraining MobileNeRF; with QEM initialization, training took an average of 8 minutes and 40 seconds.
As shown in Table~\ref{tab:synthetic}, our simplified radiance-field representations achieve slightly higher PSNR scores across most scenes.

\begin{table}
\caption{
Rendering time in the \textit{MipNeRF360}~\cite{barron2022mipnerf360} experiment.
``Full'' denotes the original image resolution.
The number of trainable parameters is included for reference.
}
\label{tab:fps_mip360}
\centering
\small
\begin{tabular}{lcccc}
\toprule
& \multicolumn{3}{c}{FPS $\uparrow$ across resolutions} \\
\cmidrule(lr){2-4}
Method & Full & 1/2 & 1/4 & \# Params \\
\midrule
3DGS~\cite{kerbl3Dgaussians} & \cellcolor{SecondCell}115 & \cellcolor{SecondCell}482 & \cellcolor{SecondCell}1.32K & 885K \\
TS~\cite{Held2025Triangle} & 88.8 & 370 & 1.00K & 885K \\
TexGS~\cite{chao2025texturedgaussians} & 16.8 & 49.1 & 94.8 & 151M \\
\textbf{Ours} (CUDA) & \cellcolor{BestCell}\textbf{1.96K} & \cellcolor{BestCell}\textbf{6.11K} & \cellcolor{BestCell}\textbf{13.7K} & 67.5M \\
\midrule
\textbf{Ours} (laptop) & 146 & 447 & 879 & 67.5M \\
\bottomrule
\end{tabular}
\end{table}

\begin{figure}
\centering
\includegraphics[width=\linewidth]{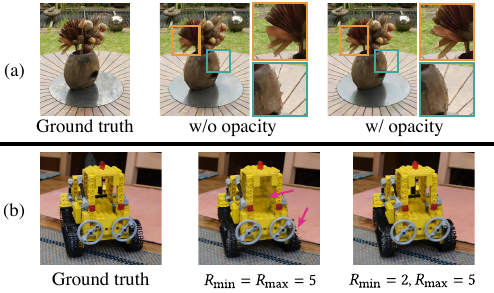}
\caption{%
Ablation studies.
(a) Without opacity learning, triangles are unable to represent fine sub-triangle boundaries.
(b) Without our multi-resolution texture scheme, the instability of color optimization leads to noisier geometry (pink arrows).
For a clearer comparison, we disable coarse-to-fine optimization in setting (b).
}
\label{fig:ablation}
\end{figure}

\paragraph*{Additional results}
Figure~\ref{fig:teaser} shows that scenes optimized using our method enable real-time rendering across diverse platforms and devices.
We implement feature lookup and opacity masking using standard vertex and fragment shaders, and perform MLP evaluation as a deferred-rendering pass in a separate shader.
Figure~\ref{fig:ablation} presents ablation studies of our key components.

\section{Conclusion}
\label{sec:conclusion}

We introduced a differentiable rasterization method for reconstructing high-fidelity 3D scenes from multi-view RGB images.
Our opaque, single-layer shading model limits the representation of transparent objects and thin structures (e.g., hair and fur), which are more naturally handled by volumetric primitives.
Because each pixel's color is determined by a single primitive rather than a blend of many, our method does not yet match state-of-the-art volumetric pipelines (e.g., AdaptiveShells~\cite{adaptiveshells2023}).
Closing this gap and achieving high-quality anti-aliasing for thin structures while retaining rendering efficiency remain important directions for future work.
We believe that our stochastic opacity masking not only enables cross-platform rendering and seamless transfer of digital 3D content but also opens new possibilities for differentiable rasterization techniques that require gradients through inherently discrete operations.

\section*{Acknowledgements}
We thank the anonymous reviewers for their valuable feedback.
This work was supported by JSPS KAKENHI Grant Number 23KJ0699
and by JST ASPIRE Grant Number JPMJAP2401.

{
    \small
    \bibliographystyle{ieeenat_fullname}
    \bibliography{main}

@String(CVPR= {IEEE Conf. Comput. Vis. Pattern Recog.})

@String(ICCV= {Int. Conf. Comput. Vis.})

@String(ECCV= {Eur. Conf. Comput. Vis.})

@String(TOG= {ACM Trans. Graph.})

@String(CVPR  = {CVPR})

@String(ICCV  = {ICCV})

@String(ECCV  = {ECCV})

@String(TOG   = {ACM TOG})

@article{kato2020differentiable,
  title={Differentiable rendering: A survey},
  author={Kato, Hiroharu and Beker, Deniz and Morariu, Mihai and Ando, Takahiro and Matsuoka, Toru and Kehl, Wadim and Gaidon, Adrien},
  journal={arXiv preprint arXiv:2006.12057},
  year={2020}
}

@article{Laine2020diffrast,
  title   = {Modular Primitives for High-Performance Differentiable Rendering},
  author  = {Samuli Laine and Janne Hellsten and Tero Karras and Yeongho Seol and Jaakko Lehtinen and Timo Aila},
  journal = {ACM Transactions on Graphics},
  year    = {2020},
  volume  = {39},
  number  = {6}
}

@inproceedings{Hasselgren2021,
  title     = {Appearance-Driven Automatic 3D Model Simplification},
  author    = {Jon Hasselgren and Jacob Munkberg and Jaakko Lehtinen and Miika Aittala and Samuli Laine},
  booktitle = {Eurographics Symposium on Rendering},
  year      = {2021}
}

@Article{kerbl3Dgaussians,
      author       = {Kerbl, Bernhard and Kopanas, Georgios and Leimk{\"u}hler, Thomas and Drettakis, George},
      title        = {3D Gaussian Splatting for Real-Time Radiance Field Rendering},
      journal      = {ACM Transactions on Graphics},
      number       = {4},
      volume       = {42},
      month        = {July},
      year         = {2023},
      url          = {https://repo-sam.inria.fr/fungraph/3d-gaussian-splatting/}
}

@inproceedings{mildenhall2020nerf,
 title={NeRF: Representing Scenes as Neural Radiance Fields for View Synthesis},
 author={Ben Mildenhall and Pratul P. Srinivasan and Matthew Tancik and Jonathan T. Barron and Ravi Ramamoorthi and Ren Ng},
 year={2020},
 booktitle={ECCV},
}

@article{liu2019softras,
  title={Soft Rasterizer: A Differentiable Renderer for Image-based 3D Reasoning},
  author={Liu, Shichen and Li, Tianye and Chen, Weikai and Li, Hao},
  journal={The IEEE International Conference on Computer Vision (ICCV)},
  month = {Oct},
  year={2019}
}

@inproceedings{Zhang2025Radiance,
author = {Zhang, Ziyi and Roussel, Nicolas and Muller, Thomas and Zeltner, Tizian and Nimier-David, Merlin and Rousselle, Fabrice and Jakob, Wenzel},
title = {Radiance Surfaces: Optimizing Surface Representations with a 5D Radiance Field Loss},
year = {2025},
isbn = {9798400715402},
publisher = {Association for Computing Machinery},
address = {New York, NY, USA},
url = {https://doi.org/10.1145/3721238.3730713},
doi = {10.1145/3721238.3730713},
booktitle = {Proceedings of the Special Interest Group on Computer Graphics and Interactive Techniques Conference Conference Papers},
articleno = {21},
numpages = {10},
location = {
},
series = {SIGGRAPH Conference Papers '25}
}

@inproceedings{pidhorskyi2024rasterized,
  title={Rasterized Edge Gradients: Handling Discontinuities Differentiably},
  author={Pidhorskyi, Stanislav and Simon, Tomas and Schwartz, Gabriel and Wen, He and Sheikh, Yaser and Saragih, Jason},
  booktitle={European Conference on Computer Vision},
  pages={335--352},
  year={2024},
  organization={Springer}
}

@article{Held2025Triangle,
  title = {Triangle Splatting for Real-Time Radiance Field Rendering},
  author = {Held, Jan and Vandeghen, Renaud and Deliege, Adrien and Hamdi, Abdullah and Cioppa, Anthony and Giancola, Silvio and Vedaldi, Andrea and Ghanem, Bernard and Tagliasacchi, Andrea and Van Droogenbroeck, Marc},
  journal = {arXiv},
  year = {2025},
}

@article{kingma2014adam,
  title={Adam: A method for stochastic optimization},
  author={Kingma, Diederik P and Ba, Jimmy},
  journal={arXiv preprint arXiv:1412.6980},
  year={2014}
}

@inproceedings{Huang2DGS2024,
    title={2D Gaussian Splatting for Geometrically Accurate Radiance Fields},
    author={Huang, Binbin and Yu, Zehao and Chen, Anpei and Geiger, Andreas and Gao, Shenghua},
    publisher = {Association for Computing Machinery},
    booktitle = {SIGGRAPH 2024 Conference Papers},
    year      = {2024},
    doi       = {10.1145/3641519.3657428}
}

@article{ravi2020accelerating,
  title={Accelerating 3d deep learning with pytorch3d},
  author={Ravi, Nikhila and Reizenstein, Jeremy and Novotny, David and Gordon, Taylor and Lo, Wan-Yen and Johnson, Justin and Gkioxari, Georgia},
  journal={arXiv preprint arXiv:2007.08501},
  year={2020}
}

@inproceedings{loper2014opendr,
  title={OpenDR: An approximate differentiable renderer},
  author={Loper, Matthew M and Black, Michael J},
  booktitle={Computer Vision--ECCV 2014: 13th European Conference, Zurich, Switzerland, September 6-12, 2014, Proceedings, Part VII 13},
  pages={154--169},
  year={2014},
  organization={Springer}
}

@article{Li:2018:DMC,
    title = {Differentiable Monte Carlo Ray Tracing through Edge Sampling},
    author = {Li, Tzu-Mao and Aittala, Miika and Durand, Fr{\'e}do and Lehtinen, Jaakko},
    journal = {ACM Trans. Graph. (Proc. SIGGRAPH Asia)},
    volume = {37},
    number = {6},
    pages = {222:1--222:11},
    year = {2018}
}

@article{NimierDavidVicini2019Mitsuba2,
    author = {Merlin Nimier-David and Delio Vicini and Tizian Zeltner and Wenzel Jakob},
    title = {Mitsuba 2: A Retargetable Forward and Inverse Renderer},
    journal = {Transactions on Graphics (Proceedings of SIGGRAPH Asia)},
    volume = {38},
    number = {6},
    year = {2019},
    month = dec,
    doi = {10.1145/3355089.3356498}
}

@article{Loubet2019Reparameterizing,
    author = {Guillaume Loubet and Nicolas Holzschuch and Wenzel Jakob},
    title = {Reparameterizing Discontinuous Integrands for Differentiable Rendering},
    journal = {Transactions on Graphics (Proceedings of SIGGRAPH Asia)},
    volume = {38},
    number = {6},
    year = {2019},
    month = dec,
    doi = {10.1145/3355089.3356510}
}

@article{Vicini2022sdf,
    author = {Delio Vicini and Sébastien Speierer and Wenzel Jakob},
    title = {Differentiable Signed Distance Function Rendering},
    journal = {Transactions on Graphics (Proceedings of SIGGRAPH)},
    volume = {41},
    number = {4},
    pages = {125:1--125:18},
    year = {2022},
    month = jul,
    doi = {10.1145/3528223.3530139}
}

@inproceedings{Bangaru2022NeuralSDFReparam,
  title = {Differentiable Rendering of Neural SDFs through Reparameterization},
  author = {Bangaru, Sai and Gharbi, Michael and Li, Tzu-Mao and Luan, Fujun and Sunkavalli, Kalyan and Hasan, Milos and Bi, Sai and Xu, Zexiang and Bernstein, Gilbert and Durand, Fredo},
  booktitle = {ACM SIGGRAPH Asia 2022 Conference Proceedings},
  articleno = {22},
  numpages = {9},
  year = {2022},
  publisher = {Association for Computing Machinery},
  address = {New York, NY, USA},
  location = {Daegu, Republic of Korea},
  series = {SIGGRAPH Asia '22},
  url = {https://doi.org/10.1145/3550469.3555397},
  doi = {10.1145/3550469.3555397}
}

@article{Nicolet2021Large,
    author = {Baptiste Nicolet and Alec Jacobson and Wenzel Jakob},
    title = {Large Steps in Inverse Rendering of Geometry},
    journal = {ACM Transactions on Graphics (Proceedings of SIGGRAPH Asia)},
    volume = {40},
    number = {6},
    year = {2021},
    month = dec,
    doi = {10.1145/3478513.3480501}
}

@article{kajiya1984ray,
  title={Ray tracing volume densities},
  author={Kajiya, James T and Von Herzen, Brian P},
  journal={ACM SIGGRAPH computer graphics},
  volume={18},
  number={3},
  pages={165--174},
  year={1984},
  publisher={ACM New York, NY, USA}
}

@InProceedings{Held20243DConvex,
title = {{3D} Convex Splatting: Radiance Field Rendering with {3D} Smooth Convexes},
author = {Held, Jan and Vandeghen, Renaud and Hamdi, Abdullah and Deliege, Adrien and Cioppa, Anthony and Giancola, Silvio and Vedaldi, Andrea and Ghanem, Bernard and Van Droogenbroeck, Marc},
booktitle = {Proceedings of the IEEE/CVF Conference on Computer Vision and Pattern Recognition (CVPR)},
year = {2025},
}

@article{wang2021neus,
  title={Neus: Learning neural implicit surfaces by volume rendering for multi-view reconstruction},
  author={Wang, Peng and Liu, Lingjie and Liu, Yuan and Theobalt, Christian and Komura, Taku and Wang, Wenping},
  journal={arXiv preprint arXiv:2106.10689},
  year={2021}
}

@inproceedings{li2023neuralangelo,
  title={Neuralangelo: High-Fidelity Neural Surface Reconstruction},
  author={Li, Zhaoshuo and M\"uller, Thomas and Evans, Alex and Taylor, Russell H and Unberath, Mathias and Liu, Ming-Yu and Lin, Chen-Hsuan},
  booktitle={IEEE Conference on Computer Vision and Pattern Recognition ({CVPR})},
  year={2023}
}

@article{yariv2021volume,
  title={Volume rendering of neural implicit surfaces},
  author={Yariv, Lior and Gu, Jiatao and Kasten, Yoni and Lipman, Yaron},
  journal={Advances in Neural Information Processing Systems},
  volume={34},
  pages={4805--4815},
  year={2021}
}

@inproceedings{decoret2003billboard,
author = {D\'{e}coret, Xavier and Durand, Fr\'{e}do and Sillion, Fran\c{c}ois X. and Dorsey, Julie},
title = {Billboard clouds for extreme model simplification},
year = {2003},
isbn = {1581137095},
publisher = {Association for Computing Machinery},
address = {New York, NY, USA},
url = {https://doi.org/10.1145/1201775.882326},
doi = {10.1145/1201775.882326},
booktitle = {ACM SIGGRAPH 2003 Papers},
pages = {689–696},
numpages = {8},
location = {San Diego, California},
series = {SIGGRAPH '03}
}

@inproceedings{chen2019dibrender,
title={Learning to Predict 3D Objects with an Interpolation-based Differentiable Renderer},
author={Wenzheng Chen and Jun Gao and Huan Ling and Edward Smith and Jaakko Lehtinen and Alec Jacobson and Sanja Fidler},
booktitle={Advances In Neural Information Processing Systems},
year={2019}
}

@article{mueller2022instant,
    author = {Thomas M\"uller and Alex Evans and Christoph Schied and Alexander Keller},
    title = {Instant Neural Graphics Primitives with a Multiresolution Hash Encoding},
    journal = {ACM Trans. Graph.},
    issue_date = {July 2022},
    volume = {41},
    number = {4},
    month = jul,
    year = {2022},
    pages = {102:1--102:15},
    articleno = {102},
    numpages = {15},
    url = {https://doi.org/10.1145/3528223.3530127},
    doi = {10.1145/3528223.3530127},
    publisher = {ACM},
    address = {New York, NY, USA},
}

@article{barron2022mipnerf360,
    title={Mip-NeRF 360: Unbounded Anti-Aliased Neural Radiance Fields},
    author={Jonathan T. Barron and Ben Mildenhall and 
            Dor Verbin and Pratul P. Srinivasan and Peter Hedman},
    journal={CVPR},
    year={2022}
}

@article{yifan2019differentiable,
  title={Differentiable surface splatting for point-based geometry processing},
  author={Yifan, Wang and Serena, Felice and Wu, Shihao and {\"O}ztireli, Cengiz and Sorkine-Hornung, Olga},
  journal={ACM Transactions On Graphics (TOG)},
  volume={38},
  number={6},
  pages={1--14},
  year={2019},
  publisher={ACM New York, NY, USA}
}

@inproceedings{chao2025texturedgaussians,
            title={Textured Gaussians for Enhanced 3D Scene Appearance Modeling},
            author={Brian Chao and Hung-Yu Tseng and Lorenzo Porzi and Chen Gao and Tuotuo Li and Qinbo Li and Ayush Saraf and Jia-Bin Huang and Johannes Kopf and Gordon Wetzstein and Changil Kim},
            year={2025},
            booktitle={CVPR}
          }

@article{zhangmanyworld,
    author = {Zhang, Ziyi and Roussel, Nicolas and Jakob, Wenzel},
    title = {Many-Worlds Inverse Rendering},
    year = {2025},
    publisher = {Association for Computing Machinery},
    address = {New York, NY, USA},
    issn = {0730-0301},
    url = {https://doi.org/10.1145/3767318},
    doi = {10.1145/3767318},
    journal = {ACM Trans. Graph.},
    month = sep
}

@InProceedings{plenoxels,
    author    = {Fridovich-Keil, Sara and Yu, Alex and Tancik, Matthew and Chen, Qinhong and Recht, Benjamin and Kanazawa, Angjoo},
    title     = {Plenoxels: Radiance Fields Without Neural Networks},
    booktitle = {Proceedings of the IEEE/CVF Conference on Computer Vision and Pattern Recognition (CVPR)},
    month     = {June},
    year      = {2022},
    pages     = {5501-5510}
}

@article{barron2021mipnerf,
    title={Mip-NeRF: A Multiscale Representation 
           for Anti-Aliasing Neural Radiance Fields},
    author={Jonathan T. Barron and Ben Mildenhall and 
            Matthew Tancik and Peter Hedman and 
            Ricardo Martin-Brualla and Pratul P. Srinivasan},
    journal={ICCV},
    year={2021}
}

@inproceedings{chen2022mobilenerf,
  title={MobileNeRF: Exploiting the Polygon Rasterization Pipeline
	for Efficient Neural Field Rendering on Mobile Architectures},
  author={Zhiqin Chen and Thomas Funkhouser
	and Peter Hedman and Andrea Tagliasacchi},
  booktitle={The Conference on Computer Vision and Pattern Recognition (CVPR)},
  year={2023}
}

@article{Held2025Triangle+,
  title = {Triangle Splatting+: Differentiable Rendering with Opaque Triangles},
  author = {Held, Jan and Vandeghen, Renaud and Son, Sanghyun and Rebain, Daniel and Gadelha, Matheus and Zhou, Yi and Lin, Ming C. and Van Droogenbroeck, Marc and Tagliasacchi, Andrea},
  journal = {arXiv},
  year = {2025},
}

@misc{xu2024texturegs,
    title={Texture-GS: Disentangling the Geometry and Texture for 3D Gaussian Splatting Editing}, 
    author={Tian-Xing Xu and Wenbo Hu and Yu-Kun Lai and Ying Shan and Song-Hai Zhang},
    year={2024},
    eprint={2403.10050},
    archivePrefix={arXiv},
    primaryClass={cs.CV}
}

@misc{xu2024SuperGaussians,
      title={SuperGaussians: Enhancing Gaussian Splatting Using Primitives with Spatially Varying Colors}, 
      author={Rui Xu and Wenyue Chen and Jiepeng Wang and Yuan Liu and Peng Wang and Lin Gao and Shiqing Xin and Taku Komura and Xin Li and Wenping Wang},
      year={2024},
      eprint={2411.18966},
      archivePrefix={arXiv},
      primaryClass={cs.CV},
      url={https://arxiv.org/abs/2411.18966}, 
}

@article{rong2024gstex,
    title={GStex: Per-primitive texturing of 2D gaussian splatting for decoupled appearance and geometry modeling},
    author={Rong, Victor and Chen, Jingxiang and Bahmani, Sherwin and Kutulakos, Kiriakos N and Lindell, David B},
    journal={arXiv preprint arXiv:2409.12954},
    year={2024}
}

@article{svitov2024billboard,
  title={BillBoard Splatting (BBSplat): Learnable Textured Primitives for Novel View Synthesis},
  author={Svitov, David and Morerio, Pietro and Agapito, Lourdes and Del Bue, Alessio},
  journal={arXiv preprint arXiv:2411.08508},
  year={2024}
}

@article{weiss2024gaussian,
  title={Gaussian billboards: Expressive 2d gaussian splatting with textures},
  author={Weiss, Sebastian and Bradley, Derek},
  journal={arXiv preprint arXiv:2412.12734},
  year={2024}
}

@inproceedings{kv2025stochasticsplats,
  title={StochasticSplats: Stochastic Rasterization for Sorting-Free 3D Gaussian Splatting},
  author={Kheradmand, Shakiba and Vicini, Delio and Kopanas, George and Lagun, Dmitry and Yi, Kwang Moo and Matthews, Mark and Tagliasacchi, Andrea},
  booktitle={Proceedings of the IEEE/CVF International Conference on Computer Vision (ICCV)},
  year={2025}
}

@inproceedings{Esposito2025VolSurfs,
  author    = {Esposito, Stefano and Chen, Anpei and Reiser, Christian and Rota Bulò, Samuel and Porzi, Lorenzo and Schwarz, Katja and Richardt, Christian and Zollhoefer, Michael and Kontschieder, Peter and Geiger, Andreas},
  title     = {Volumetric Surfaces: Representing Fuzzy Geometries with Layered Meshes},
  booktitle = {IEEE/CVF Conference on Computer Vision and Pattern Recognition (CVPR)},
  year      = {2025}
}

@article{bengio2013estimating,
  title={Estimating or propagating gradients through stochastic neurons for conditional computation},
  author={Bengio, Yoshua and L{\'e}onard, Nicholas and Courville, Aaron},
  journal={arXiv preprint arXiv:1308.3432},
  year={2013}
}

@InProceedings{Mehta_2023_ICCV,
    author    = {Mehta, Ishit and Chandraker, Manmohan and Ramamoorthi, Ravi},
    title     = {A Theory of Topological Derivatives for Inverse Rendering of Geometry},
    booktitle = {Proceedings of the IEEE/CVF International Conference on Computer Vision (ICCV)},
    month     = {October},
    year      = {2023},
    pages     = {419-429}
}

@article{williams1992simple,
  title={Simple statistical gradient-following algorithms for connectionist reinforcement learning},
  author={Williams, Ronald J},
  journal={Machine learning},
  volume={8},
  number={3},
  pages={229--256},
  year={1992},
  publisher={Springer}
}

@article{glynn1990likelihood,
  title={Likelihood ratio gradient estimation for stochastic systems},
  author={Glynn, Peter W},
  journal={Communications of the ACM},
  volume={33},
  number={10},
  pages={75--84},
  year={1990},
  publisher={ACM New York, NY, USA}
}

@article{Weier2025PracticalInverse,
  title   = {Practical Inverse Rendering of Textured and Translucent Appearance},
  author  = {Philippe Weier and J{\'e}r{\'e}my Riviere and Ruslan Guseinov and Stephan Garbin and Philipp Slusallek and Bernd Bickel and Thabo Beeler and Delio Vicini},
  year    = {2025},
  month   = aug,
  journal = {ACM Transactions on Graphics (Proceedings of SIGGRAPH)},
  volume  = {44},
  number  = {4},
  doi     = {10.1145/3730855}
}

@InProceedings{takikawa2021nglod,
    author    = {Takikawa, Towaki and Litalien, Joey and Yin, Kangxue and Kreis, Karsten and Loop, Charles and Nowrouzezahrai, Derek and Jacobson, Alec and McGuire, Morgan and Fidler, Sanja},
    title     = {Neural Geometric Level of Detail: Real-Time Rendering With Implicit 3D Shapes},
    booktitle = {Proceedings of the IEEE/CVF Conference on Computer Vision and Pattern Recognition (CVPR)},
    month     = {June},
    year      = {2021},
    pages     = {11358-11367}
}

@article{yuksel2010mesh,
  title={Mesh colors},
  author={Yuksel, Cem and Keyser, John and House, Donald H},
  journal={ACM Transactions on Graphics (TOG)},
  volume={29},
  number={2},
  pages={1--11},
  year={2010},
  publisher={ACM New York, NY, USA}
}

@inproceedings{yuksel2017mesh,
author = {Yuksel, Cem},
title = {Mesh color textures},
year = {2017},
isbn = {9781450351010},
publisher = {Association for Computing Machinery},
address = {New York, NY, USA},
url = {https://doi.org/10.1145/3105762.3105780},
doi = {10.1145/3105762.3105780},
booktitle = {Proceedings of High Performance Graphics},
articleno = {17},
numpages = {11},
location = {Los Angeles, California},
series = {HPG '17}
}

@article{thies2019deferred,
  title={Deferred neural rendering: Image synthesis using neural textures},
  author={Thies, Justus and Zollh{\"o}fer, Michael and Nie{\ss}ner, Matthias},
  journal={Acm Transactions on Graphics (TOG)},
  volume={38},
  number={4},
  pages={1--12},
  year={2019},
  publisher={ACM New York, NY, USA}
}

@InProceedings{worchel:2022:nds,
      author    = {Worchel, Markus and Diaz, Rodrigo and Hu, Weiwen and Schreer, Oliver and Feldmann, Ingo and Eisert, Peter},
      title     = {Multi-View Mesh Reconstruction with Neural Deferred Shading},
      booktitle = {Proceedings of the IEEE/CVF Conference on Computer Vision and Pattern Recognition (CVPR)},
      month     = {June},
      year      = {2022},
      pages     = {6187-6197}
}

@article{ling2022vectoradam,
  title={Vectoradam for rotation equivariant geometry optimization},
  author={Ling, Selena Zihan and Sharp, Nicholas and Jacobson, Alec},
  journal={Advances in Neural Information Processing Systems},
  volume={35},
  pages={4111--4122},
  year={2022}
}

@article{paszke2019pytorch,
  title={Pytorch: An imperative style, high-performance deep learning library},
  author={Paszke, Adam and Gross, Sam and Massa, Francisco and Lerer, Adam and Bradbury, James and Chanan, Gregory and Killeen, Trevor and Lin, Zeming and Gimelshein, Natalia and Antiga, Luca and others},
  journal={Advances in neural information processing systems},
  volume={32},
  year={2019}
}

@article{wang2004image,
  title={Image quality assessment: from error visibility to structural similarity},
  author={Wang, Zhou and Bovik, Alan C and Sheikh, Hamid R and Simoncelli, Eero P},
  journal={IEEE transactions on image processing},
  volume={13},
  number={4},
  pages={600--612},
  year={2004},
  publisher={IEEE}
}

@misc{nanobind,
   author = {Wenzel Jakob},
   year = {2022},
   note = {https://github.com/wjakob/nanobind},
   title = {nanobind: tiny and efficient C++/Python bindings}
}

@article{adaptiveshells2023,
    author = {Zian Wang and Tianchang Shen and Merlin Nimier-David and Nicholas Sharp and Jun Gao 
        and Alexander Keller and Sanja Fidler and Thomas M\"uller and Zan Gojcic},
    title = {Adaptive Shells for Efficient Neural Radiance Field Rendering},
    journal = {ACM Trans. Graph.},
    issue_date = {December 2023},
    volume = {42},
    number = {6},
    year = {2023},
    articleno = {259},
    numpages = {15},
    url = {https://doi.org/10.1145/3618390},
    doi = {10.1145/3618390},
    publisher = {ACM},
    address = {New York, NY, USA}
  }

@inproceedings{garland1997,
author = {Garland, Michael and Heckbert, Paul S.},
title = {Surface simplification using quadric error metrics},
year = {1997},
isbn = {0897918967},
publisher = {ACM Press/Addison-Wesley Publishing Co.},
address = {USA},
url = {https://doi.org/10.1145/258734.258849},
doi = {10.1145/258734.258849},
booktitle = {Proceedings of the 24th Annual Conference on Computer Graphics and Interactive Techniques},
pages = {209–216},
numpages = {8},
series = {SIGGRAPH '97}
}

@article{eldar1997farthest,
  title={The farthest point strategy for progressive image sampling},
  author={Eldar, Yuval and Lindenbaum, Michael and Porat, Moshe and Zeevi, Yehoshua Y},
  journal={IEEE transactions on image processing},
  volume={6},
  number={9},
  pages={1305--1315},
  year={1997},
  publisher={IEEE}
}

@inproceedings{glorot2010understanding,
  title={Understanding the difficulty of training deep feedforward neural networks},
  author={Glorot, Xavier and Bengio, Yoshua},
  booktitle={Proceedings of the thirteenth international conference on artificial intelligence and statistics},
  pages={249--256},
  year={2010},
  organization={JMLR Workshop and Conference Proceedings}
}

@inproceedings{he2016deep,
  title={Deep residual learning for image recognition},
  author={He, Kaiming and Zhang, Xiangyu and Ren, Shaoqing and Sun, Jian},
  booktitle={Proceedings of the IEEE conference on computer vision and pattern recognition},
  pages={770--778},
  year={2016}
}

@inproceedings{kheradmand20243d,
    title = {3D Gaussian Splatting as Markov Chain Monte Carlo},
    author = {Kheradmand, Shakiba and Rebain, Daniel and Sharma, Gopal and Sun, Weiwei and Tseng, Yang-Che and Isack, Hossam and Kar, Abhishek and Tagliasacchi, Andrea and Yi, Kwang Moo},
    booktitle = {Advances in Neural Information Processing Systems (NeurIPS)},
    year = {2024},
    note = {Spotlight Presentation},
   }
}

% =============================================================
% Supplementary Material — formatting preamble
% =============================================================
\clearpage
\setcounter{section}{0}
\setcounter{figure}{0}
\setcounter{table}{0}
\setcounter{equation}{0}
\setcounter{algocf}{0}
\renewcommand{\thesection}{\Alph{section}}
\renewcommand{\thefigure}{A\arabic{figure}}
\renewcommand{\thetable}{A\arabic{table}}
\renewcommand{\theequation}{A\arabic{equation}}
\renewcommand{\thealgocf}{A\arabic{algocf}}

\section*{Supplementary Material for DiffSoup}
\label{app:supp}

This supplementary material provides an accompanying video and a detailed document to complement the main paper.
The video presents the optimization sequence, novel-view synthesis results with geometry visualizations, and a real-time demonstration in which we interactively render a scene containing \emph{all} models from the \textit{NeRF-Synthetic}~\cite{mildenhall2020nerf} and \textit{Shelly}~\cite{adaptiveshells2023} datasets on a mobile device (iPhone 15).
This mobile demonstration achieves an average of 51~FPS at $2\text{K}\times1\text{K}$ resolution on camera views computed using Fibonacci spiral sampling over the upper hemisphere (included in our code release).
In addition, the following sections provide further details on our experimental setup, method, mathematical derivations, and additional figures and tables.

\section{Experimental Setup}
\label{sec:setup}

\subsection{Implementation}
We implement our method in Python using PyTorch~\cite{paszke2019pytorch} for automatic differentiation.
Our differentiable rasterizer is implemented in CUDA and exposed as a custom PyTorch operator via nanobind~\cite{nanobind}.
Edge splitting routines are likewise implemented in C++ and invoked directly from Python.
Our implementation is publicly available at \url{https://github.com/kenji-tojo/diffsoup}.

\subsection{Hardware}
Our experiments were conducted primarily on a workstation equipped with an Intel Core i9-14900K CPU, 64 GB of RAM, and an NVIDIA GeForce RTX 4090 GPU with 24 GB of VRAM.
In particular, all results produced by our method were obtained in a single-GPU setting on the RTX 4090.
To parallelize the evaluation of baseline methods, we additionally used a separate machine equipped with an NVIDIA RTX A6000 GPU with 48 GB of VRAM.
This secondary hardware was used solely to accelerate baseline training; all reported results are from single-GPU runs, and all methods fit within the memory budget of the RTX 4090.

The original MobileNeRF results~\cite{chen2022mobilenerf} were obtained using eight NVIDIA V100 GPUs. Reproducing this multi-GPU configuration using a single GPU would not be feasible with our hardware and would also be disproportionate relative to the other baselines.
For context, all other baselines—and our method—typically complete training in roughly 30 minutes on a single GPU. In contrast, MobileNeRF's training is considerably more computationally demanding: keeping all hyperparameters fixed and running on a single GPU results in training times of around 6 hours.
Replicating their original eight-GPU setup would multiply the computational cost and exceed our available resources.
To enable a fair comparison, we therefore run the MobileNeRF code on the same single GPU used for all other methods. This results in an effective batch size one eighth of the original, while the number of iterations and the learning rate remain identical to those reported in the paper.

\begin{figure*}[t]
\centering
\includegraphics[width=\linewidth]{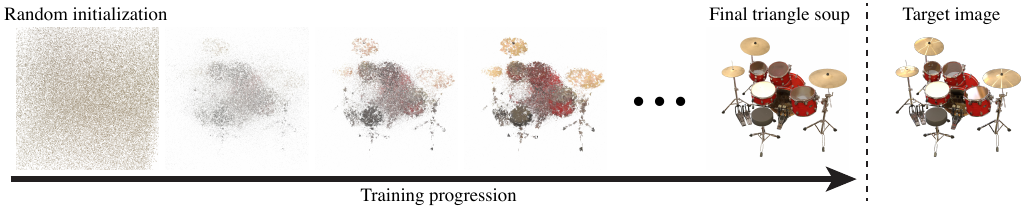}
\caption{%
(Left) Starting from a randomly initialized triangle soup, training gradually drives irrelevant triangles to become transparent, and these are regularly removed by our adaptive resampling step.
This process yields a clean geometry that captures even thin structures and challenging topology.
(Right) The corresponding ground-truth image is shown for reference.
}
\label{fig:random}
\end{figure*}

\section{Method Details}
\label{sec:training}
This section provides additional details about our method.

\subsection{Stochastic Opacity Masking Pseudocode}

\begin{algorithm}[t]
\caption{Stochastic opacity masking}
\label{alg:stochastic}
\DontPrintSemicolon
\SetAlgoNoEnd\SetAlgoNoLine

\KwIn{fragments $\mathcal{F} =\{\mathbf{f}_1,\,\dots,\,\mathbf{f}_{\vert\mathcal{F}\vert}\}$}
\KwOut{pixel color $\hat{\mathbf{C}}$ and color/opacity gradients}

\BlankLine
\textbf{1. Forward pass}\;
compute $f^{*}$ and $\hat{\mathbf{C}}$ by sampling thresholds $\boldsymbol{\tau}$, and store $(\mathbf{f}_{f^{*}}, \hat{\mathbf{C}})$ in image-sized buffers\;

\BlankLine
\textbf{2. Backward pass}\;

\textbf{(a) Color gradient:} differentiate $\mathcal{L}_1(\hat{\mathbf{C}})$\;

\textbf{(b) Opacity gradient:}\;
\ForEach{$f' \in \{1,\,\dots,\,\vert\mathcal{F}\vert\}$}{
  \lIf{$f' = f^{*}$}{$s \leftarrow 1/\alpha_{f^{*}}$}
  \lElseIf{$D_{f'} < D_{f^{*}}$}{$s \leftarrow -1/(1-\alpha_{f'})$}
  \lElse{$s \leftarrow 0$}
  add $\mathcal{L}_1(\hat{\mathbf{C}})\, s$ to the opacity gradient of $f'$\;
}
\end{algorithm}

Algorithm~\ref{alg:stochastic} summarizes the forward and backward passes of our stochastic opacity masking procedure, as described in the main paper.

\subsection{Hyperparameters}
For Adam~\cite{kingma2014adam} and VectorAdam~\cite{ling2022vectoradam}, we set $\beta_1 = 0.9$, $\beta_2 = 0.999$, and $\epsilon = 1.0\times10^{-8}$.
We exponentially anneal the learning rate so that it is reduced by a factor of 100 over the course of training.
We use an initial learning rate of $5.0\times10^{-2}$ for color and alpha features, and $1.0\times10^{-2}$ for neural network weights.
For vertex positions, we set the initial learning rate to $1.0\times10^{-2}$ for real scenes and $1.0\times10^{-3}$ for synthetic scenes, as geometry in synthetic datasets tends to be more compact.

\subsection{MLP Architecture}
For deferred shading, we apply a shared-weight multilayer perceptron (MLP) with two 16-dimensional hidden layers to the per-pixel feature vector.
As input to the MLP, we concatenate the 7-dimensional color feature vector with the per-pixel view direction encoded using spherical harmonic bases up to degree~2, resulting in a $7 + 9 = 16$-dimensional input vector.
All MLP weights are initialized following Xavier initialization~\cite{glorot2010understanding}.
All texture feature vectors are initialized to zero, which results in an initial value of 0.5 after applying the sigmoid.

We observe that color vibrancy can be diminished after passing through the MLP.
To mitigate this, we interpret the first four channels of the color feature as RGBA and compute the final RGB output as a blend of this RGBA feature and the MLP prediction, yielding a parameter-free, lightweight skip-connection layer~\cite{he2016deep}.

\subsection{Coarse-to-Fine Optimization}
As described in the main paper, we transition from the coarse feature grid to the fine grid at iteration~5000.
This is performed by sampling the coarse color feature at each grid coordinate and copying it to the new $R_{\mathrm{min}}$ level.
Because coarse-stage training often produces oversaturated alpha values that hinder later shape refinement, we reinitialize the alpha feature during this transition.

\subsection{Random Triangle Soup Initialization}
For the synthetic data experiments, we additionally report results obtained using a random initialization of the triangle-soup geometry.
We generate this initialization by uniformly sampling seed points inside the bounding box used for MobileNeRF~\cite{chen2022mobilenerf} and placing a triangle around each point with a random orientation and a radius of~0.01 (the bounding-box scale typically ranges from~2 to~4).

For a randomly initialized triangle soup, we apply a custom adaptive triangle-resampling procedure to avoid unnecessary triangles.
Specifically, we first remove triangles that are fully transparent—that is, those that receive no pixel coverage in any training image.
We then iteratively split the longest edges until the triangle count reaches the target value, ensuring a consistent number of triangles throughout training.
Figure~\ref{fig:random} shows the optimization progression from random initialization.

\subsection{Data Size}
Our representation uses 9 floating-point numbers (3 per vertex) to encode the geometry of each triangle.
This matches the data footprint of standard 3DGS variants~\cite{kerbl3Dgaussians}, which also require 9 floating-point values (3 for the center, 3 for the rotation, and 3 for the scale) per primitive.

We use a triangle texture resolution of $R_{\mathrm{max}} = 5$ at inference time, which produces $(2^4 + 1)\times(2^5 + 1) = 561$ grid points per triangle.
At this resolution, all triangle textures can be packed into a 4K grid provided the number of triangles remains below
\begin{equation}
    4096^2 / 561 \approx 29{,}905.
\end{equation}
Note that 4K textures are widely supported across rendering environments, including mobile devices.

In the comparison on the \textit{MipNeRF360}~\cite{barron2022mipnerf360} dataset, Textured Gaussians (TexGS)~\cite{chao2025texturedgaussians} employ a $50\times50$ texture grid (2500 texels) per primitive.
Because their representation combines RGBA textures with per-primitive spherical harmonic (SH) coefficients, the effective number of texture channels per primitive is $2500\times4 = 10{,}000$, whereas our method uses only $561\times8 = 4{,}488$ channels per triangle (recall that we use a 7-channel color feature plus alpha).

\begin{figure*}[t]
\centering
\includegraphics[width=\linewidth]{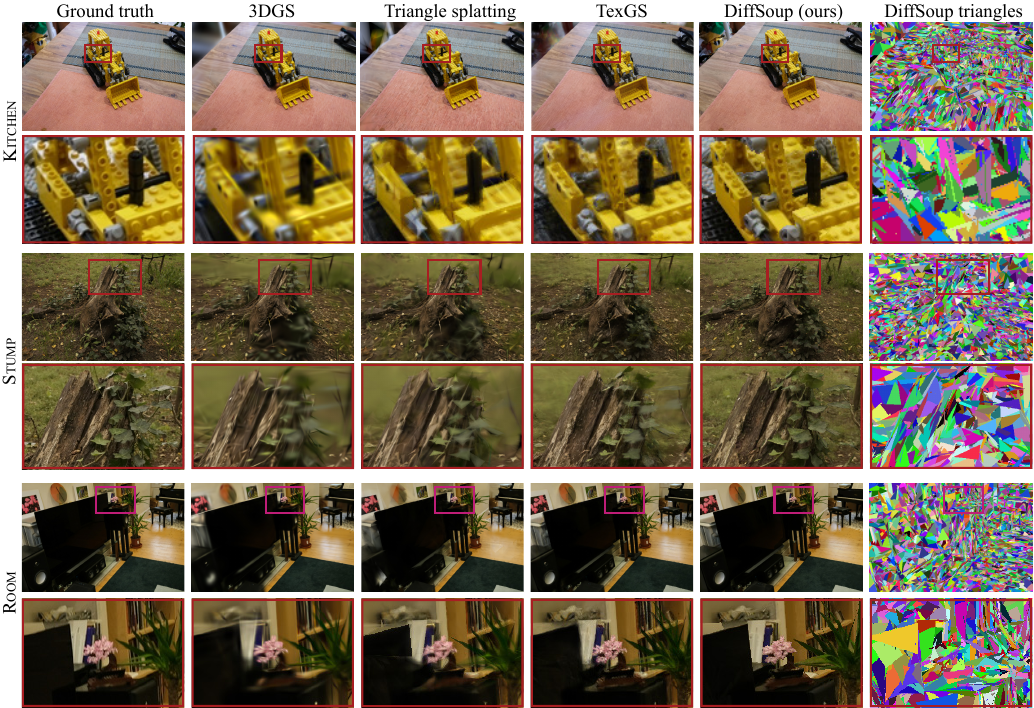}
\caption{%
Additional qualitative results on the \textit{MipNeRF360}~\cite{barron2022mipnerf360} dataset.
(Left to right) Ground-truth test images; views synthesized by 3DGS~\cite{kerbl3Dgaussians}, Triangle Splatting~\cite{Held2025Triangle}, TexGS~\cite{chao2025texturedgaussians}, our method; and a visualization of our reconstructed triangle geometry.
Rows correspond to different scenes, and all models use 15K primitives.
}
\label{fig:mip360_extra}
\end{figure*}

\section{Mathematical Derivations}
\label{sec:proofs}
In this section, we present the full mathematical derivations corresponding to the equations introduced in the method section of the main paper.

\subsection{Equivalence of stochastic opacity masking and radiance field loss}

In the main paper, we claim that our stochastic opacity masking technique selects a fragment $f$ with probability
\begin{equation}
\label{suppl:eq:stochastic}
    p(f) =
    \left(
    \prod_{f'\in\{f';\,D_{f'}<D_{f}\}}(1-\alpha_{f'})
    \right)
    \,\alpha_{f},
\end{equation}
which coincides with the sample weight $w_f$ used as the blending weight in the radiance field loss~\cite{Zhang2025Radiance} and in the standard volume rendering formulation of NeRFs~\cite{mildenhall2020nerf}.
This equivalence follows directly from the observation that fragment $f$ is selected as the foremost visible fragment if and only if all fragments in front of it are \emph{not} visible and fragment $f$ itself is visible.
Fragments behind $f$ do not influence its visibility and are therefore marginalized out of the probability computation, yielding the expression in~\eqref{suppl:eq:stochastic}.

\subsection{Application of the likelihood-ratio identity}

The main paper derives the unbiased estimator using the likelihood-ratio identity~\cite{glynn1990likelihood,williams1992simple}.
Although this identity is classical and not part of our technical contribution, we briefly verify that it applies to our gradient expression as well, to keep the presentation self-contained.

We first express the expected value in summation form:
\begin{equation}
    E_{p(f)}[\mathcal{L}(\hat{\mathcal{C}})]
    =
    \sum_{f\in\mathcal{F}}
    \mathcal{L}(\hat{\mathcal{C}})\,p(f).
\end{equation}
By differentiating the expression with respect to $\alpha_{f'}$, we obtain
\begin{equation}
\begin{aligned}
    \frac{\partial}{\partial\alpha_{f'}}
    E_{p(f)}[\mathcal{L}(\hat{\mathbf{C}})]
    &=
    \sum_{f\in\mathcal{F}}
    \frac{\partial\mathcal{L}(\hat{\mathcal{C}})}{\partial\alpha_{f'}}\,p(f) \\
    &+
    \sum_{f\in\mathcal{F}}
    \mathcal{L}(\hat{\mathcal{C}})\frac{\partial{p(f)}}{\partial\alpha_{f'}},
\end{aligned}
\end{equation}
where the first term is simply the expected value
\begin{equation}
    \sum_{f\in\mathcal{F}}
    \frac{\partial\mathcal{L}(\hat{\mathcal{C}})}{\partial\alpha_{f'}}\,p(f)
    =
    E_{p(f)}
    \left[
    \frac{\partial\mathcal{L}(\hat{\mathcal{C}})}{\partial\alpha_{f'}}
    \right],
\end{equation}
by the definition of expectation.
The second term can also be written as an expectation by observing that
\begin{equation}
\begin{aligned}
    \sum_{f\in\mathcal{F}}
    \mathcal{L}(\hat{\mathcal{C}})\frac{\partial{p(f)}}{\partial\alpha_{f'}}
    &=
    \sum_{f\in\mathcal{F}}
    \mathcal{L}(\hat{\mathcal{C}})
    \left(
    \frac{1}{p(f)}
    \frac{\partial{p(f)}}{\partial\alpha_{f'}}
    \right)\,
    p(f)\\
    &=
    \sum_{f\in\mathcal{F}}
    \mathcal{L}(\hat{\mathcal{C}})
    \frac{\partial{\log{p(f)}}}{\partial\alpha_{f'}}
    \,
    p(f)\\
    &=
    E_{p(f)}
    \left[
    \mathcal{L}(\hat{\mathcal{C}})
    \,
    \frac{\partial{\log{p(f)}}}{\partial\alpha_{f'}}
    \right].
\end{aligned}
\end{equation}
Combining the two parts yields the gradient estimator introduced in the main paper.

\subsection{Gradient of the score term}

Based on the probability $p(f)$, the main paper derives the score term as
\begin{equation}
\frac{\partial\log p(f)}{\partial\alpha_{f'}}
=
\begin{cases}
    1/\alpha_f, & f'=f,\\
    -1/(1-\alpha_{f'}), & D_{f'} < D_f,
\end{cases}
\end{equation}
which is used for computing the loss gradient.
This expression follows directly from applying the chain rule,
\begin{equation}
\frac{\partial\log p(f)}{\partial\alpha_{f'}}
=
\frac{1}{p(f)}
\frac{\partial{p(f)}}{\partial\alpha_{f'}},
\end{equation}
where we observe that
\begin{equation}
\frac{\partial{p(f)}}{\partial\alpha_{f'}}
=
\begin{cases}
    p(f)/\alpha_f, & f'=f,\\
    -p(f)/(1-\alpha_{f'}), & D_{f'} < D_f,
\end{cases}
\end{equation}
which is obtained by differentiating the product expression of $p(f)$ with respect to $\alpha_{f'}$.

\definecolor{BestCell}{HTML}{DFF0D8}
\definecolor{SecondCell}{HTML}{D9EDF7}

\begin{table*}[t]
\caption{%
Additional quantitative comparisons with MobileNeRF~\cite{chen2022mobilenerf} on synthetic scenes from the \textit{NeRF-Synthetic}~\cite{mildenhall2020nerf} and \textit{Shelly}~\cite{adaptiveshells2023} datasets.
In addition to the PSNR values reported in the main paper, we report SSIM (top table) and LPIPS (bottom table) scores for each scene. The face count for MobileNeRF is averaged across scenes, whereas all other methods use the exact face count specified.
See the main paper for further discussion.
}
\label{tab:synthetic_extra}
\centering
\small
\begin{tabular}{lccccccccc}
  \toprule
  Method & \multicolumn{5}{c}{\textit{NeRF-Synthetic}~\cite{mildenhall2020nerf} (SSIM $\uparrow$)} & \multicolumn{4}{c}{\textit{Shelly}~\cite{adaptiveshells2023} (SSIM $\uparrow$)} \\
  \cmidrule(lr){2-6}
  \cmidrule(lr){7-10}
   & \textsc{ship} & \textsc{chair} & \textsc{mic} & \textsc{drums} & \# Faces $\downarrow$ & \textsc{khady} & \textsc{kitten} & \textsc{pug} & \# Faces $\downarrow$ \\
  \midrule
  MobileNeRF~\cite{chen2022mobilenerf} & 0.817 & 0.958 & 0.961 & 0.910 & 159K & 0.810 & 0.931 & 0.857 & 275K \\
  MobileNeRF + QEM~\cite{garland1997} & 0.554 & 0.793 & 0.857 & 0.739 & 15K & 0.686 & 0.847 & 0.729 & 15K \\
  \midrule
  Ours w/ QEM init. & \cellcolor{BestCell}\textbf{0.848} & \cellcolor{BestCell}\textbf{0.975} & \cellcolor{BestCell}\textbf{0.975} & \cellcolor{BestCell}\textbf{0.931} & 15K & \cellcolor{BestCell}\textbf{0.838} & \cellcolor{BestCell}\textbf{0.951} & \cellcolor{BestCell}\textbf{0.903} & 15K \\
  Ours w/ random init. & \cellcolor{SecondCell}0.828 & \cellcolor{SecondCell}0.972 & \cellcolor{SecondCell}0.972 & \cellcolor{SecondCell}0.927 & 15K & \cellcolor{SecondCell}0.838 & \cellcolor{SecondCell}0.950 & \cellcolor{SecondCell}0.899 & 15K \\
  \bottomrule
\end{tabular}
\begin{tabular}{lccccccccc}
  \toprule
  Method & \multicolumn{5}{c}{\textit{NeRF-Synthetic}~\cite{mildenhall2020nerf} (LPIPS $\downarrow$)} & \multicolumn{4}{c}{\textit{Shelly}~\cite{adaptiveshells2023} (LPIPS $\downarrow$)} \\
  \cmidrule(lr){2-6}
  \cmidrule(lr){7-10}
   & \textsc{ship} & \textsc{chair} & \textsc{mic} & \textsc{drums} & \# Faces $\downarrow$ & \textsc{khady} & \textsc{kitten} & \textsc{pug} & \# Faces $\downarrow$ \\
  \midrule
  MobileNeRF~\cite{chen2022mobilenerf} & 0.173 & 0.042 & 0.073 & 0.094 & 159K & 0.205 & 0.099 & 0.192 & 275K \\
  MobileNeRF + QEM~\cite{garland1997} & 0.453 & 0.227 & 0.235 & 0.314 & 15K & 0.306 & 0.170 & 0.294 & 15K \\
  \midrule
  Ours w/ QEM init. & \cellcolor{BestCell}\textbf{0.119} & \cellcolor{BestCell}\textbf{0.024} & \cellcolor{BestCell}\textbf{0.037} & \cellcolor{BestCell}\textbf{0.066} & 15K & \cellcolor{BestCell}\textbf{0.153} & \cellcolor{BestCell}\textbf{0.074} & \cellcolor{BestCell}\textbf{0.116} & 15K \\
  Ours w/ random init. & \cellcolor{SecondCell}0.147 & \cellcolor{SecondCell}0.027 & \cellcolor{SecondCell}0.043 & \cellcolor{SecondCell}0.071 & 15K & \cellcolor{SecondCell}0.154 & \cellcolor{SecondCell}0.077 & \cellcolor{SecondCell}0.120 & 15K \\
  \bottomrule
\end{tabular}
\end{table*}

\begin{table}
\caption{%
Quantitative results for the ablation study on the \textit{MipNeRF360}~\cite{barron2022mipnerf360} dataset.
From top to bottom: (1) our method with opacity fixed to 1, (2) a variant that uses only the highest-resolution feature grid with $R_{\mathrm{max}} = 5$ (i.e., without multi-resolution features and without the coarse-to-fine strategy), (3) a variant that uses multi-resolution features but does not use the coarse-to-fine strategy, and (4) our full method combining both multi-resolution features and the coarse-to-fine strategy.
PSNR, SSIM, and LPIPS scores are reported as scene-level averages.
The number of triangles is fixed at $\lvert\mathcal{T}\rvert = 15\mathrm{K}$ for all methods.
}
\label{tab:mip360_ablation}
\centering
\small
\begin{tabular}{lccc}
\toprule
Method & PSNR $\uparrow$ & SSIM $\uparrow$ & LPIPS $\downarrow$ \\
\midrule
w/o opacity & \cellcolor{SecondCell}24.65 & 0.744 & 0.205 \\ 
$R_{\mathrm{min}}=R_{\mathrm{max}}=5$ & 23.54 & 0.719 & 0.226 \\ 
$R_{\mathrm{min}}=2,\,R_{\mathrm{max}}=5$ & 24.48 & \cellcolor{SecondCell}0.747 & \cellcolor{BestCell}\textbf{0.203} \\ 
\textbf{Ours full} & \cellcolor{BestCell}\textbf{24.76} & \cellcolor{BestCell}\textbf{0.748} & \cellcolor{SecondCell}0.204 \\ 
\bottomrule
\end{tabular}
\end{table}

\begin{figure}[b]
\centering
\includegraphics[width=\linewidth]{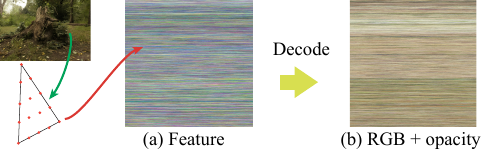}
\caption{%
We represent RGB color and opacity within each triangle by linearly interpolating values stored at subdivision points.
(a) Each subdivision point stores eight 32-bit floating-point values: a 7-dimensional color feature vector and one opacity value.
(b) The color feature vector is decoded into RGB using a small neural network, while the final channel is used directly as opacity.
}
\label{fig:texture_vis}
\end{figure}

\section{Additional Figures}
Figure~\ref{fig:mip360_extra} shows additional qualitative comparison results on the \textit{MipNeRF360}~\cite{barron2022mipnerf360} dataset.
Overall, our opaque textured-triangle representation produces the sharpest visuals under a tight primitive budget, and in particular yields noticeably sharper results than TexGS~\cite{chao2025texturedgaussians}, which uses textured translucent primitives.
Figure~\ref{fig:texture_vis} visualizes the texture memory used in our method.

\begin{figure}[t]
\centering
\includegraphics[width=\linewidth]{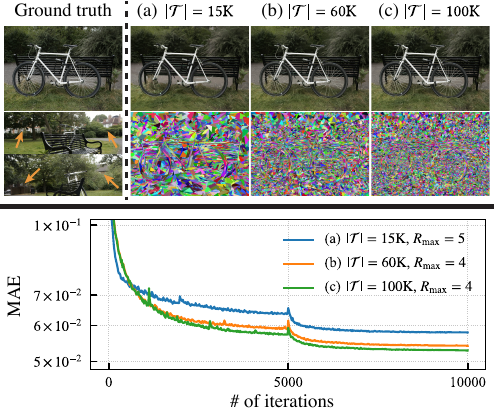}
\caption{%
(Top row, left to right) Ground-truth images for the \textsc{Bicycle} scene from the \textit{MipNeRF360}~\cite{barron2022mipnerf360} dataset.
This scene exhibits an extremely large spatial extent, as indicated by the distant objects highlighted with orange arrows.
(a) Triangle-soup rendering and reconstructed geometry using our method with $\lvert\mathcal{T}\rvert = 15\mathrm{K}$ triangles and a texture resolution of $R_{\mathrm{max}} = 5$.
(b) Rendering and geometry using $\lvert\mathcal{T}\rvert = 60\mathrm{K}$ triangles and $R_{\mathrm{max}} = 4$.
(c) Rendering and geometry using $\lvert\mathcal{T}\rvert = 100\mathrm{K}$ triangles and $R_{\mathrm{max}} = 4$.
(Bottom row) Test error measured as mean absolute error (MAE) over training iterations for the three configurations, demonstrating convergence behavior and clarifying the representational capacity.
}
\label{fig:convergence}
\end{figure}

\paragraph*{Challenging case}
Outdoor scenes are generally more challenging to represent with a small number of triangles than indoor or synthetic scenes, primarily due to their extremely large spatial extent.
A particularly notable example in our experiments is the \textsc{Bicycle} scene from the \textit{MipNeRF360}~\cite{barron2022mipnerf360} dataset (Figure~\ref{fig:convergence}).
In this scene, important visual details—such as the spokes of the bicycle wheel—occupy only a tiny region within a very large overall scene volume.

Our adaptive resampling method encourages roughly uniform pixel coverage per triangle to make the most effective use of the available texture resolution.
In such extreme cases, however, geometric structures that fall below a certain relative world-space scale can be missed. This effect can be seen in Figure~\ref{fig:convergence}~(a), where large objects are reconstructed clearly under a tight triangle budget, but the thin wheel spokes are lost.

A potential remedy is to decrease the average screen-space footprint of each triangle by increasing the number of triangles.
Figure~\ref{fig:convergence}~(b) illustrates this approach: we halve the screen-space edge-split threshold, increase the target triangle count to four times that of (a), and reduce the texture resolution by one subdivision level.
This adjustment yields noticeably better reconstruction of fine geometry while keeping the overall texture memory roughly unchanged.
If the budget allows, further fidelity can be achieved by reducing the edge-split threshold to one third of (a) and increasing the triangle budget to 100K, as shown in Figure~\ref{fig:convergence}~(c).

While uniformly increasing the number of triangles improves representational capacity, as seen in the convergence plot of Figure~\ref{fig:convergence}, it is still highly desirable to remain within the original tight triangle budget.
We believe this is achievable through more sophisticated strategies for allocating triangle density and texture resolution—beyond our current uniform, image-space-driven resampling—which could further improve the reconstruction of challenging outdoor scenes under very limited primitive budgets (e.g., the 15K budget used in our experiments).
Designing such spatially adaptive edge-splitting and texture-resolution schemes remains an important direction for future work.

\begin{figure*}[t]
\centering
\includegraphics[width=\linewidth]{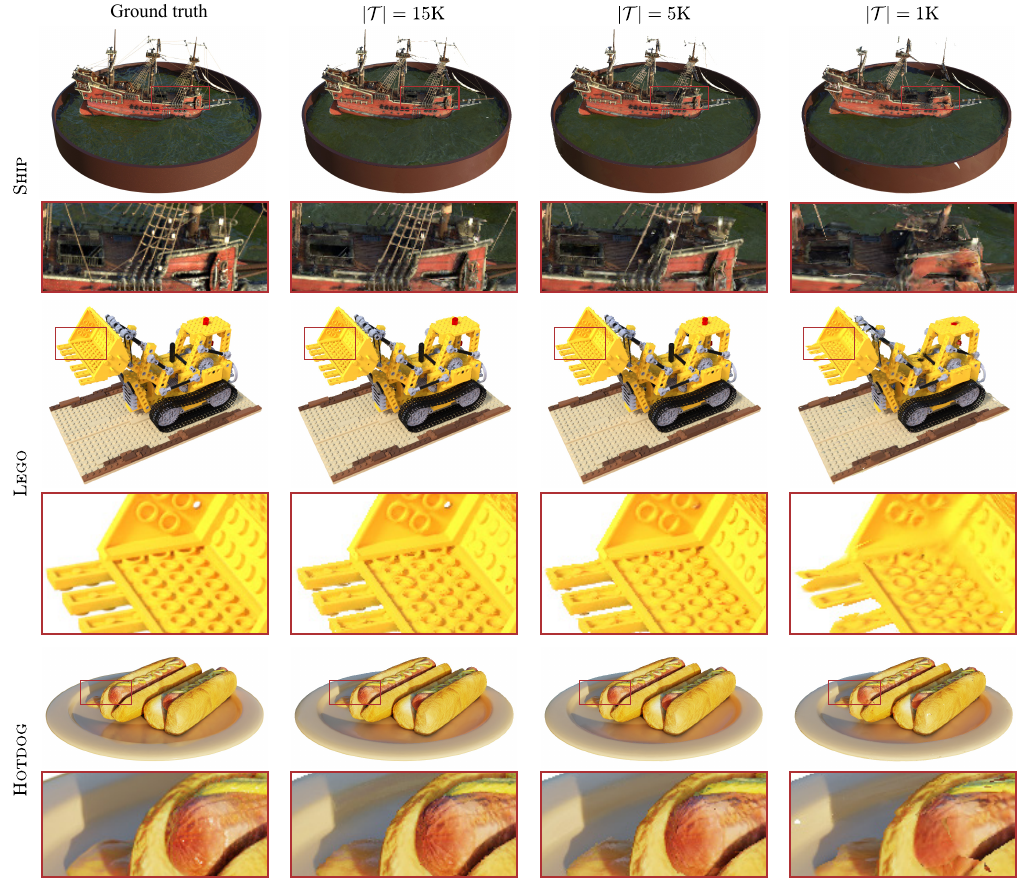}
\caption{%
Effect of varying the triangle budget on reconstruction quality
for \textit{NeRF-Synthetic}~\cite{mildenhall2020nerf} scenes.
Each row shows (left to right) a ground-truth view followed
by our results with
$\vert\mathcal{T}\vert = 15\mathrm{K}$, $5\mathrm{K}$,
and $1\mathrm{K}$ triangles, respectively.
}
\label{fig:budget_sweep}
\end{figure*}

\section{Additional Quantitative Results}
This section presents additional quantitative results for experiments presented in the main paper.
Table~\ref{tab:synthetic_extra} reports SSIM and LPIPS scores for the comparison against MobileNeRF~\cite{chen2022mobilenerf} on synthetic datasets, supplementing the PSNR results presented in the main paper.
In terms of these metrics, the two variants of our method consistently achieve the best and second-best performance.

Table~\ref{tab:mip360_ablation} reports additional quantitative scores for the ablation study conducted in the main paper.
The inability to represent sub-triangle geometry when binary opacity is not used is especially evident in the degraded SSIM and LPIPS scores.
Interestingly, the variant that uses the multi-resolution feature grid throughout optimization (i.e., without the coarse-to-fine strategy) provides a competitive alternative, whereas the variant that relies solely on the highest-resolution grid consistently performs the worst.
This highlights the effectiveness of our multi-resolution feature-grid design.
Our full method additionally incorporates the coarse-to-fine strategy and consistently achieves the best or second-best results across all metrics.
We observe that coarse-to-fine training often improves triangle coverage and geometric fitting, as reflected in higher PSNR scores compared to the multi-resolution-only variant.

\section{Additional Comparisons and Ablations}
\label{sec:additional_analyses}

This section presents additional experiments and analyses that complement the main paper, including a comparison with 3DGS-MCMC~\cite{kheradmand20243d}, storage costs, training time, and a primitive-budget sweep.

\subsection{Comparison with 3DGS-MCMC}
3DGS-MCMC~\cite{kheradmand20243d} provides explicit control over the primitive count while relying on the same primitives as 3DGS, making it a relevant additional baseline.
We evaluate it under the same 15K budget on the \textit{MipNeRF360}~\cite{barron2022mipnerf360} dataset.
Table~\ref{tab:mcmc} reports quality metrics, and Table~\ref{tab:mcmc_fps} reports rendering speed.
Although 3DGS-MCMC outperforms vanilla 3DGS in reconstruction quality, it remains slower than our method and lower in quality than the textured approaches (TexGS and ours) in this low-budget regime.

\subsection{Storage Cost}
Table~\ref{tab:storage} reports the inference-time GPU memory footprint and storage cost of the models used in our experiments.
Among the textured methods (i.e., TexGS, MobileNeRF, and ours), our approach achieves the lowest memory and storage cost.
Both MobileNeRF and our method leverage 8-bit RGBA textures, which reduce memory usage and enable efficient storage and data transmission via PNG compression.
We were unable to load the scene shown in our supplementary video, comprising 14 models, on a mobile device using MobileNeRF; after loading 6 models, the total memory footprint reached 1.58~GB, exceeding the browser memory limit.
In contrast, our full scene requires only 916~MB.

\begin{table}
\caption{%
Reconstruction quality comparison with 3DGS-MCMC~\cite{kheradmand20243d} on the \textit{MipNeRF360}~\cite{barron2022mipnerf360} dataset.
All methods use 15K primitives.
}
\label{tab:mcmc}
\centering
\small
\begin{tabular}{lccc}
\toprule
Method & PSNR $\uparrow$ & SSIM $\uparrow$ & LPIPS $\downarrow$ \\
\midrule
3DGS~\cite{kerbl3Dgaussians} & 23.72 & 0.664 & 0.420 \\
3DGS-MCMC~\cite{kheradmand20243d} & 24.22 & 0.692 & 0.389 \\
TexGS~\cite{chao2025texturedgaussians} & \cellcolor{BestCell}\textbf{24.80} & \cellcolor{SecondCell}0.697 & \cellcolor{SecondCell}0.270 \\
\textbf{Ours} & \cellcolor{SecondCell}24.76 & \cellcolor{BestCell}\textbf{0.748} & \cellcolor{BestCell}\textbf{0.204} \\
\bottomrule
\end{tabular}
\end{table}

\begin{table}
\caption{%
Rendering speed comparison with 3DGS-MCMC~\cite{kheradmand20243d} on the \textit{MipNeRF360}~\cite{barron2022mipnerf360} dataset.
All methods use 15K primitives.
}
\label{tab:mcmc_fps}
\centering
\small
\begin{tabular}{lccc}
\toprule
& \multicolumn{3}{c}{FPS $\uparrow$ across resolutions} \\
\cmidrule(lr){2-4}
Method & Full & 1/2 & 1/4 \\
\midrule
3DGS~\cite{kerbl3Dgaussians} & 115 & 482 & \cellcolor{SecondCell}1.32K \\
3DGS-MCMC~\cite{kheradmand20243d} & \cellcolor{SecondCell}214 & \cellcolor{SecondCell}598 & 943 \\
TexGS~\cite{chao2025texturedgaussians} & 16.8 & 49.1 & 94.8 \\
\textbf{Ours} (CUDA) & \cellcolor{BestCell}\textbf{1.96K} & \cellcolor{BestCell}\textbf{6.11K} & \cellcolor{BestCell}\textbf{13.7K} \\
\bottomrule
\end{tabular}
\end{table}

\begin{table}
\caption{%
Average GPU memory footprint and storage cost of the models used in our experiments.
The average primitive count is also reported.
}
\label{tab:storage}
\centering
\small
\begin{tabular}{lccc}
\toprule
Method & Memory $\downarrow$ & Storage $\downarrow$ & \# Prims \\
\midrule
3DGS~\cite{kerbl3Dgaussians} & \cellcolor{BestCell}\textbf{3.38MB} & \cellcolor{BestCell}\textbf{3.38MB} & 15K \\
TexGS~\cite{chao2025texturedgaussians} & 576MB & 576MB & 15K \\
MobileNeRF~\cite{chen2022mobilenerf} & 329MB & 118MB & 209K \\
\textbf{Ours} & \cellcolor{SecondCell}65.9MB & \cellcolor{SecondCell}44.3MB & 15K \\
\bottomrule
\end{tabular}
\end{table}

\begin{table}
\caption{%
Reconstruction quality (PSNR) across triangle budgets on
\textit{NeRF-Synthetic}~\cite{mildenhall2020nerf} scenes.
Quality degrades gracefully from
$\vert\mathcal{T}\vert{=}15\mathrm{K}$ to $5\mathrm{K}$,
with a more noticeable drop at $1\mathrm{K}$.
}
\label{tab:budget_sweep}
\centering
\small
\begin{tabular}{lccccc}
\toprule
& \multicolumn{5}{c}{PSNR $\uparrow$} \\
\cmidrule(lr){2-6}
\# Triangles & \textsc{Ship} & \textsc{Mic} & \textsc{Lego} & \textsc{Hotdog} & \textsc{Ficus} \\
\midrule
$\vert\mathcal{T}\vert=15\mathrm{K}$ & 26.68 & 30.18 & 29.78 & 33.53 & 27.51 \\
$\vert\mathcal{T}\vert=5\mathrm{K}$ & 26.12 & 30.03 & 28.21 & 33.45 & 27.39 \\
$\vert\mathcal{T}\vert=1\mathrm{K}$ & 24.34 & 28.26 & 24.36 & 31.40 & 24.27 \\
\bottomrule
\end{tabular}
\end{table}

\subsection{Training Time}
For the \textsc{Garden} scene at the $1/4$ resolution reported in Table~\ref{tab:mip360} of the main paper, 3DGS completes training in 4.5 minutes.
Our method requires a longer training time of 13.5 minutes, reflecting the use of our prototype software rasterizer during training.
Despite this, our training is still significantly faster than TexGS, which also relies on textures and requires 46 minutes.
Importantly, our stochastic opacity masking introduces only a small overhead: ablating it reduces the training time to 11.5 minutes.

\subsection{Triangle Budget Analysis}
Increasing the number of primitives improves reconstruction fidelity, especially for outdoor scenes with large spatial extent (see Figure~\ref{fig:convergence}).
For per-object scenes in the \textit{NeRF-Synthetic}~\cite{mildenhall2020nerf} dataset, the triangle count can be further reduced with only minor quality degradation.
We evaluate this on a subset of scenes using the same QEM initialization as in our MobileNeRF comparison; Table~\ref{tab:budget_sweep} and Figure~\ref{fig:budget_sweep} report quantitative and qualitative results, respectively.
For example, on \textsc{Ship}, lowering the budget from $\vert\mathcal{T}\vert = 15\mathrm{K}$ to $5\mathrm{K}$ decreases PSNR by only 0.56\,dB, with the main visible artifact being a slight loss of thin structures.
For scenes with an overall compact shape, such as \textsc{Hotdog}, the drop is even smaller and barely perceptible.
In contrast, reducing to $1\mathrm{K}$ triangles leads to noticeable degradation across all scenes, both qualitatively and quantitatively.
Automatically selecting the optimal budget given a user-specified error tolerance is an interesting direction for future work.

\end{document}